\documentclass[twocolumn,showpacs,preprintnumbers,amsmath,amssymb,nofootinbib,prc,superscriptaddress]{revtex4-1}

\usepackage{graphicx}
\usepackage{dcolumn}
\usepackage{bm}
\usepackage{multirow}

\usepackage{amssymb,amsmath,txfonts}
\usepackage{maybemath}
\usepackage[pdftex]{xcolor}
\usepackage{hyperref}

\def\apj{Astroph. J.~}                                                    
\def\apjl{Astroph. J.Lett.~}                                                    
\def\aap{Astron.\&Astrophys.~}
\def\epja{Eur. Phys. J. A~}                                               
\def\nima{Nucl. Instr. Meth. A~}                                          
\def\nimb{Nucl. Instr. Meth. B~}                                          
\def\np{Nucl. Phys.}                                                         
\def\npa{Nucl. Phys. A~}                                                  
\def\pra{Phys. Rev. A~}                                                   
\def\prc{Phys. Rev. C~}                                                   
\def\plb{Phys. Lett. B~}                                                  
\def\zpa{Z. Phys. A~}                                                     

\newcommand{\elem}[2]{{}^{#2}\mathrm{#1}}
\newcommand{\fiNag}{^{15}\mathrm{N}(\alpha,\gamma)^{19}\mathrm{F}}

\begin{document}
\title{Measurement of 1323 and 1487\,keV resonances in ${^{15}\mathrm{N}(\alpha,\gamma)^{19}\mathrm{F}}$ with the recoil separator ERNA}
\author{A.~\surname{Di Leva}}\email[Corresponding author: ]{antonino.dileva@unina.it}\affiliation{Dipartimento di Fisica ``E. Pancini", Universit\`a di Napoli ``Federico II'', Napoli, Italy}\affiliation{INFN, Sezione di Napoli, Napoli, Italy} 
\author{G.~\surname{Imbriani}}\affiliation{Dipartimento di Fisica ``E. Pancini", Universit\`a di Napoli ``Federico II'', Napoli, Italy}\affiliation{INFN, Sezione di Napoli, Napoli, Italy}
\author{R.~\surname{Buompane}}\affiliation{Dipartimento di Matematica e Fisica, Universit\`a degli Studi della Campania ``L. Vanvitelli'', Caserta, Italy}\affiliation{INFN, Sezione di Napoli, Napoli, Italy}
\author{L.~\surname{Gialanella}}\affiliation{Dipartimento di Matematica e Fisica, Universit\`a degli Studi della Campania ``L. Vanvitelli'', Caserta, Italy}\affiliation{INFN, Sezione di Napoli, Napoli, Italy}
\author{A.~\surname{Best}}\affiliation{Dipartimento di Fisica ``E. Pancini", Universit\`a di Napoli ``Federico II'', Napoli, Italy}\affiliation{INFN, Sezione di Napoli, Napoli, Italy}
\author{S.~\surname{Cristallo}}\affiliation{INAF, Osservatorio Astronomico di Collurania, Teramo, Italy}\affiliation{INFN, Sezione di Perugia, Perugia, Italy}
\author{M.~\surname{De Cesare}}\affiliation{Dipartimento di Metodologie e Tecnologie per le Osservazioni e Misure, Centro Italiano Ricerche Aerospaziali, Capua, Italy}\affiliation{Dipartimento di Matematica e Fisica, Universit\`a degli Studi della Campania ``L. Vanvitelli'', Caserta, Italy}\affiliation{INFN, Sezione di Napoli, Napoli, Italy}
\author{A.~\surname{D'Onofrio}}\affiliation{Dipartimento di Matematica e Fisica, Universit\`a degli Studi della Campania ``L. Vanvitelli'', Caserta, Italy}\affiliation{INFN, Sezione di Napoli, Napoli, Italy}
\author{J.~G.~\surname{Duarte}}\affiliation{Dipartimento di Matematica e Fisica, Universit\`a degli Studi della Campania ``L. Vanvitelli'', Caserta, Italy}\affiliation{INFN, Sezione di Napoli, Napoli, Italy}
\author{L.~R.~\surname{Gasques}}\affiliation{Departamento de F\'isica Nuclear, Instituto de F\'isica da Universidade de S\~ao Paulo, S\~ao Paulo, SP, Brazil}\affiliation{Dipartimento di Matematica e Fisica, Universit\`a degli Studi della Campania ``L. Vanvitelli'', Caserta, Italy}\affiliation{INFN, Sezione di Napoli, Napoli, Italy}
\author{L.~\surname{Morales-Gallegos}}\affiliation{SUPA, School of Physics and Astronomy, University of Edinburgh, Edinburgh, UK}\affiliation{INFN, Sezione di Napoli, Napoli, Italy}
\author{A.~\surname{Pezzella}}\affiliation{Dipartimento di Scienze Chimiche, Universit\`a di Napoli ``Federico II'', Napoli, Italy}\affiliation{INFN, Sezione di Napoli, Napoli, Italy}
\author{G.~\surname{Porzio}}\affiliation{Dipartimento di Matematica e Fisica, Universit\`a degli Studi della Campania ``L. Vanvitelli'', Caserta, Italy}\affiliation{INFN, Sezione di Napoli, Napoli, Italy}
\author{D.~\surname{Rapagnani}}\affiliation{Dipartimento di Fisica e Geologia, Universit\`a degli Studi di Perugia, Perugia, Italy}\affiliation{INFN, Sezione di Perugia, Perugia, Italy}
\author{V.~\surname{Roca}}\affiliation{Dipartimento di Fisica ``E. Pancini", Universit\`a di Napoli ``Federico II'', Napoli, Italy}\affiliation{INFN, Sezione di Napoli, Napoli, Italy}
\author{M.~\surname{Romoli}}\affiliation{INFN, Sezione di Napoli, Napoli, Italy}
\author{D.~\surname{Sch\"urmann}}\affiliation{Dipartimento di Fisica ``E. Pancini", Universit\`a di Napoli ``Federico II'', Napoli, Italy}\affiliation{INFN, Sezione di Napoli, Napoli, Italy}
\author{O.~\surname{Straniero}}\affiliation{INAF, Osservatorio Astronomico di Collurania, Teramo, Italy}\affiliation{INFN, Sezione di Napoli, Napoli, Italy}
\author{F.~\surname{Terrasi}}\affiliation{Dipartimento di Matematica e Fisica, Universit\`a degli Studi della Campania ``L. Vanvitelli'', Caserta, Italy}\affiliation{INFN, Sezione di Napoli, Napoli, Italy}
\collaboration{ERNA Collaboration}\thanks{The Authors want to commemorate M. Romano and N. De Cesare who, since its beginning, gave an invaluable contribution to the ERNA project. Their early demise is an irreplaceable loss for our Collaboration, and for the Experimental Nuclear Physics community.}

\begin{abstract}
\begin{description}
\item[Background] The origin of fluorine is a widely debated issue. Nevertheless, the $^{15}\mathrm{N}(\alpha,\gamma)^{19}\mathrm{F}$ reaction is a common feature among the various production channels so far proposed. Its reaction rate at relevant temperatures is determined by a number of narrow resonances together with  the DC component and the tails of the two broad resonances at $E_{\rm c.m.}=1323$ and 1487\,keV.
\item[Method] Measurement through the direct detection of the $^{19}\mathrm{F}$ recoil ions with the European Recoil separator for Nuclear Astrophysics (ERNA) were performed. The reaction was initiated by a $^{15}\mathrm{N}$ beam impinging onto a $^4\mathrm{He}$ windowless gas target. The observed yield of the resonances at $E_{\rm c.m.}=1323$ and 1487\,keV is used to determine their widths in the $\alpha$ and $\gamma$ channels.
\item[Results] We show that a direct measurement of the cross section of the $^{15}\mathrm{N}(\alpha,\gamma)^{19}\mathrm{F}$ reaction can be successfully obtained with the Recoil Separator ERNA, and the widths $\Gamma_\gamma$ and $\Gamma_\alpha$ of the two broad resonances have been determined. While a fair agreement is found with earlier determination of the widths of the 1487\,keV resonance, a significant difference is found for the 1323\,keV resonance $\Gamma_\alpha$.
\item[Conclusions] The revision of the widths of the two more relevant broad resonances in the $^{15}\mathrm{N}(\alpha,\gamma)^{19}\mathrm{F}$ reaction presented in this work is the first step toward a more firm determination of the reaction rate. At present, the residual uncertainty at the  temperatures of the $^{19}\mathrm{F}$ stellar nucleosynthesis is dominated by the uncertainties affecting the Direct Capture component and the 364\,keV narrow resonance, both so far investigated only through indirect experiments.
\end{description}
\end{abstract}

\pacs{26.20-f, 26.20.Fj, 23.20.Lv}

\maketitle
\section{Introduction}
\label{intro}
The origin of $^{19}\mathrm{F}$ is a widely debated issue in astrophysics. Several stellar environments have been proposed as F production sites: core-collapse Supernovae \citep{Woosley1988}, Wolf-Rayet stars \citep{Meynet2000}, and Asymptotic Giant Branch (AGB) stars \citep{Forestini1992}. Among them, only in AGB stars fluorine synthesis is confirmed  by direct spectroscopic observation of [F/Fe] enhancements, see \cite[and references therein]{Jorissen1992, Abia2009}, and recent studies seem to exclude the first two scenarios \cite{Federman2005,Palacios2005}.

It was early recognised that the $^{15}\mathrm{N}(\alpha,\gamma)^{19}\mathrm{F}$ reaction is a leading process for the $^{19}\mathrm{F}$ production when He-burning is active. Although the H burning ashes are heavily depleted in $^{15}\mathrm{N}$, which is efficiently destroyed by proton capture, these ashes are enriched in $^{14}\mathrm{N}$. Various  reaction chains may lead to the production of $^{15}\mathrm{N}$ nuclei at relatively low temperatures, $\sim100\,$MK. A likely reaction chain is $^{14}\mathrm{N}(\rm n,p)^{14}\mathrm{C}(\alpha,\gamma)^{18}\mathrm{O}(\rm p,\alpha)^{15}\mathrm{N}$, which however requires an efficient neutron source. Some $^{15}\mathrm{N}$ may be also produced by the $^{14}\mathrm{N}(\alpha,\gamma)^{18}\mathrm{F}(\beta^+)^{18}\mathrm{O}(\rm p,\alpha)^{15}\mathrm{N}$, where the protons need to be simultaneously released by the $^{14}\rm N(n,p)^{14}C$ reaction. Therefore the presence of a neutron source is a key requirement. This condition is actually fulfilled in low-mass AGB stars undergoing thermal pulses, where the $^{13}\mathrm{C}(\alpha,\rm n)^{16}\mathrm{O}$ reaction is known to be the main neutron source powering the $s$-process nucleosynthesis in their He-rich mantel \cite{Straniero1995}. The competition with some reactions that destroy $^{15}\mathrm{N}$ and/or $^{19}\mathrm{F}$, such as $^{15}$N(p,$\alpha$)$^{12}$C, $^{19}$F(n,$\gamma$)$^{20}$F, $^{19}$F(p,$\alpha$)$^{16}$O, and $^{19}$F($\alpha$,p)$^{22}$Ne, should also be carefully considered, see e.g. Refs.\cite{Imbriani2012,Lombardo2015} for recent experimental works, and Ref. \cite{Cristallo2014} for a review.
\begin{figure*}[!htb]
\begin{center}
\resizebox{.9\hsize}{!}{\includegraphics{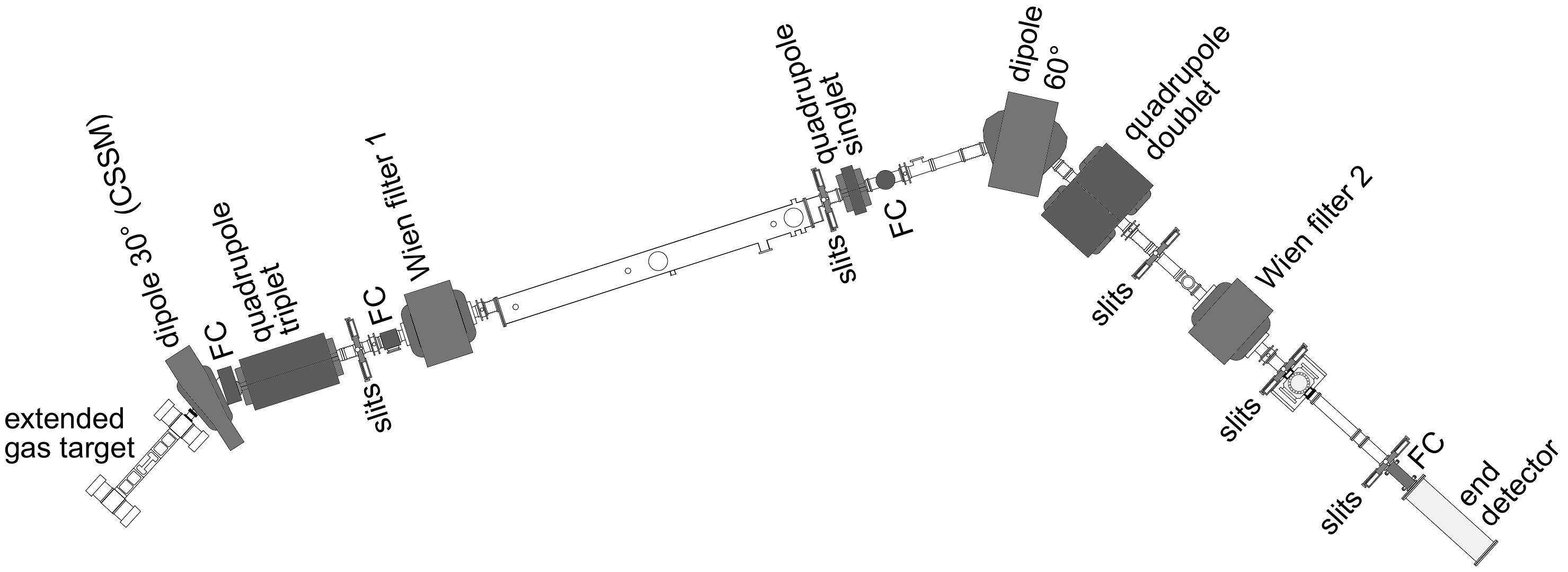}}
\caption{Schematic view of the ERNA recoil separator.}
\label{fig:ERNA}
\end{center}
\end{figure*}

The rate of the $^{15}$N($\alpha$,$\gamma$)$^{19}$F reaction at relevant AGB temperatures is determined by a number of narrow resonances, the most important being the $E_{\rm c.m.}=364\,$keV one, together with the Direct Capture (DC) component and the tails of two broad resonances at $E_{\rm c.m.}=1323$ and 1487\,keV. The strength of the $E_{\rm c.m.}=364\,$keV resonance has been determined through an indirect measurement reported in \cite{deOliveira1996}. Due to the model dependence of the result an uncertainty of a factor of 2 is assumed for this quantity. In the same work the spectroscopic factors of most of the $^{19}\rm F$ bound states were determined, and on the basis of a single particle transition model the DC component has been estimated. On this latter quantity, according to the survey in Ref. \cite{Longland2010}, an uncertainty of 40\% is generally assumed. The mentioned uncertainties influence the determination of the reaction rate at relevant AGB temperatures.

\section{Experimental setup and procedures}

The measurement of the $\fiNag$ reaction yield was performed in inverse kinematics, i.e.  a $^{15}\mathrm{N}$ beam \cite{DiLeva2012} impinging onto a $^4\mathrm{He}$ windowless gas target, using the European Recoil separator for Nuclear Astrophysics (ERNA). ERNA was originally installed and commissioned at the Dynamitron Tandem Laboratorium of the Ruhr-Universit\"at Bochum, Germany \cite{Rogalla2003,Gialanella2004,Schuermann2004}. In 2009 it was moved to the Center for Isotopic Research on Cultural and Environmental heritage (CIRCE) laboratory in Caserta, Italy \cite{Terrasi2007}.  The separator underwent a major upgrade with the addition of the Charge State Selection dipole Magnet (CSSM) directly downstream of the target. A schematic view of the present ERNA layout is shown in Fig. \ref{fig:ERNA}. The ion beam emerging from the 3\,MV tandem accelerator is transported through the CIRCE AMS beamline: a $90^\circ$ analyzing magnet and an electrostatic analizer provide the necessary ion beam purification from recoil-like contaminants. The magnetic field of the analyzing magnet is used to determine the beam energy, while its uncertainty is determined by the opening of the magnet's image slits. The settings used in the presented measurements result in a beam energy uncertainty of about 7\,keV \cite{Buompane2016}. The beam is guided into the $40^\circ$ beam line of ERNA by a switching magnet. A quadrupole triplet after the switching magnet is used to focus the beam onto the windowless gas target \cite{Schuermann2013}.
After the gas target, the separator consists sequentially of the following elements: a dipole magnet (CSSM) a quadrupole triplet (QT), a Wien filter (WF1), a quadrupole singlet (QS), a $60^\circ$ dipole magnet, a quadrupole doublet (QD), a Wien filter (WF2), and a detector for recoil identification and counting. Finally, several Faraday cups (FC), and slit systems are installed along the beam line for diagnostic purposes.
A Si detector is placed at about $25^\circ$ in the laboratory frame with respect to beam axis, and is collimated with a $\phi=1$\,mm diameter aperture in the second downstream pumping stage of the gas target. This is used to monitor the scattering rate of $^{15}\mathrm{N}$ ions on the post-stripper Ar gas, see below, needed to determine the number of projectiles impinging on the target, $N_p$. The scattering on Ar ensures a smooth behaviour of the elastic scattering yield. Calibration measurements are performed several times between the cross section measurements.

The reaction yield is given by:
\begin{equation}
  Y_i = N_p\Phi_qT_{RMS}\eta\int^{E_{\elem{N}{15}}}_{E_{\elem{N}{15}}-T_t}\frac {\sigma(E)}{\varepsilon(E)}\,dE\enspace,
  \label{eq:Yield}
\end{equation}
where $\Phi_q$ is the probability of recoils in the $q+$ charge state to enter the separator, $T_{RMS}$ is the separator transmission of recoils in charge state $q+$ to the end detector, $\eta$ is the detection efficiency, $E_{\elem{N}{15}}$ is the beam energy, $T_t$ is the target thickness, $\varepsilon(E)$ is the stopping power of N ions in He. All of these quantities have to be determined in order to extract the cross section $\sigma$ from the observed yield.

\subsection{\maybebm{\elem{He}{4}} target characterisation}
\label{sec:target}

\begin{figure}[!b]
\begin{center}
\resizebox{.9\hsize}{!}{\includegraphics{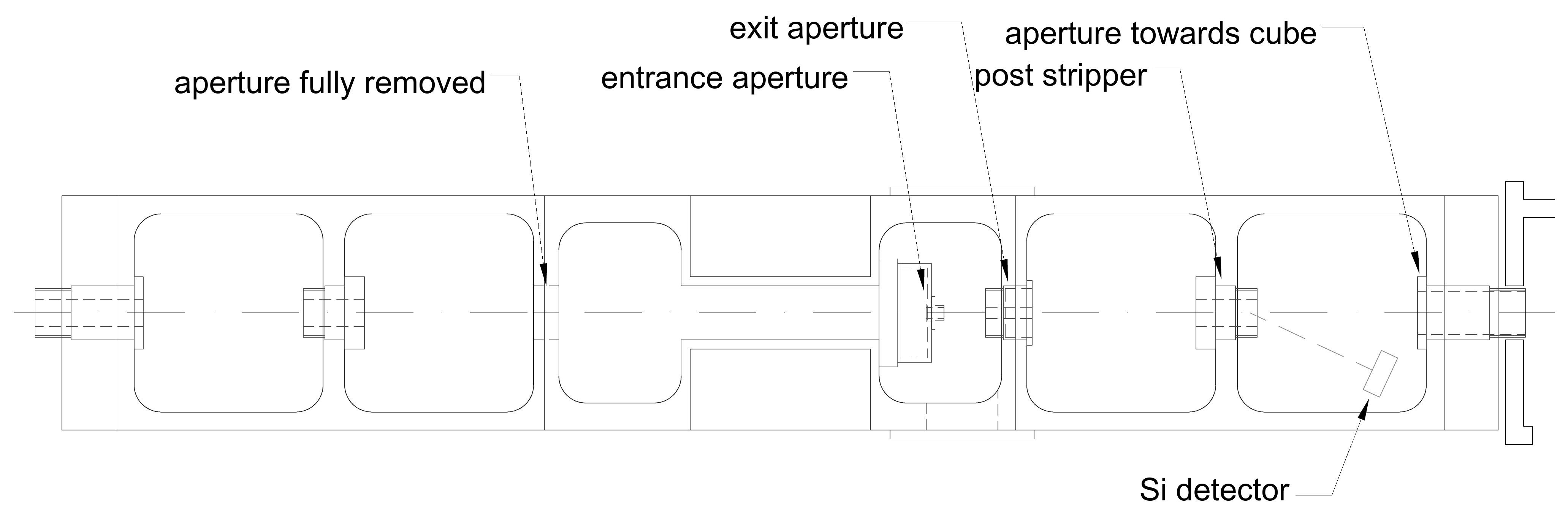}}
\caption{Schematic top view of the modifications to the extended gas target chamber. The relevant parts discussed in the text are indicated, for further details see Ref. \cite{Schuermann2013}.}
\label{fig:TargetChamber}
\end{center}
\end{figure}

The recoil separator ERNA, in order to measure the $\elem{Be}{7}(p,\gamma)\elem{B}{8}$, has been recently provided with a windowless differentially pumped $\rm H_2$ extended gas target cell \cite{Schuermann2013} with an effective length of about 300\,mm. This cell is too long to achieve the necessary angular acceptance for the measurement of the $\elem{N}{14,15}(\alpha,\gamma)\elem{F}{18,19}$ reaction cross sections. Therefore the central target cell was sectioned with a wall and appropriate apertures, as schematically shown in figure \ref{fig:TargetChamber}.
As reported in \cite{Schuermann2013}, in the aperture between the first and the second downstream pumping stages, Ar gas is injected in order to have an additional gas layer (post-stripper) that allows recoil ions to reach charge state equilibrium regardless of their actual reaction coordinates within the target.
In order to reach the needed angular acceptance, see Sec. \ref{sec:acceptance}, the downstream apertures have the following diameters: post-stripper aperture has $\phi=15$\,mm, aperture toward the downstream cube and the aperture between the two cube pumping stages $24$\,mm and $27$\,mm, respectively. 

\subsubsection{Target thickness}

We have determined the total target thickness through the measurement of the energy loss of several ions, see Table \ref{tab:CSSMscans}. The uncertainties are due to the $\Delta B$ determination, and to the uncertainty on the stopping power values. The total thickness value is $(0.54\pm0.03)\times10^{18}\rm\,atoms/cm^2$.
\begin{table}[!b]
\begin{center}
\small
\begin{tabular}{ccccccc}
\hline
\hline
 & $E_{\rm Lab}$ & $\Delta$B($\elem{He}4$) & $\varepsilon(\elem{He}4)$ & $\Delta$E($\elem{He}4$) & Thickness \\
Ion & [MeV] & [mT] & [keV\,cm$^{2}$/1E18] & [keV] & [1E18/cm$^{2}$] \\
$\elem{C}{12}$ & 3.5 & 6.13 & 64.3 & $43.3\pm3.1$ & $0.67\pm0.11$ \\
$\elem{N}{14}$ & 3.0 & 7.72 & 79.0 & $46.7\pm2.5$ & $0.59\pm0.09$ \\
$\elem{N}{15}$ & 6.3 & 3.39 & 85.0 & $43.3\pm4.8$ & $0.51\pm0.10$ \\
$\elem{O}{16}$ & 4.5 & 5.05 & 89.8 & $52.6\pm3.5$ & $0.59\pm0.08$ \\
$\elem{F}{19}$ & 4.8 & 5.09 & 103 & $50.2\pm2.7$ & $0.49\pm0.06$ \\
$\elem{F}{19}$ & 3.5 & 5.85 & 94.7 & $49.5\pm3.0$ & $0.52\pm0.07$ \\
\hline
\hline
\end{tabular}
\end{center}
\caption{Measured values, results and relevant quantities used in the target thickness determination.}
\label{tab:CSSMscans}
\end{table}%

\noindent It is worth noting that there are some issues regarding the stopping power values of N ions in He gas. This is particularly relevant since the stopping power value at resonance energy is needed to calculate the strength of a resonance from the reaction yield. In general there are not many experimental data available for gaseous targets, see e.g. \cite{IAEAwebsite}, however the stopping power of N in He was measured a significant number of times. The SRIM2003 tables appear to have a worse agreement to the experimental data with respect to the older Ziegler's 1996 calculations \cite{IAEAwebsite}, therefore stopping power values of N in He according to this latter calculation have been used in this work. The stopping power of N in He in the energy range used in the present work is essentially determined by the data of Ref. \cite{Price1993}, where a 2.5\% systematic uncertainty is reported. However since the Ziegler's 1996 is not an actual fit to the experimental data a more conservative 5\% uncertainty is assumed.

Thickness of the post-stripper alone, needed to estimate the effect on angular straggling of the recoils, was measured at the working pressure of 10\,mbar, using a 2.5\,MeV $\elem{F}{19}^{2+}$ beam. The observed shift in CSSM field is $\Delta B=(3.96\pm0.08)\,$mT, for a reference field of 1057.3\,mT. SRIM2003 tables report for F in Ar a stopping power of 412\,keV/(1E18\,atoms/cm$^2$) at this energy,  thus the corresponding thickness is  $(4.5\pm0.5)\times10^{16}\rm\,atoms/cm^2$. Most of the uncertainty is due to an assumed 10\% error on the stopping power.

\subsubsection{Density profile}

\begin{figure}[!t]
\begin{center}
\resizebox{.9\hsize}{!}{\includegraphics{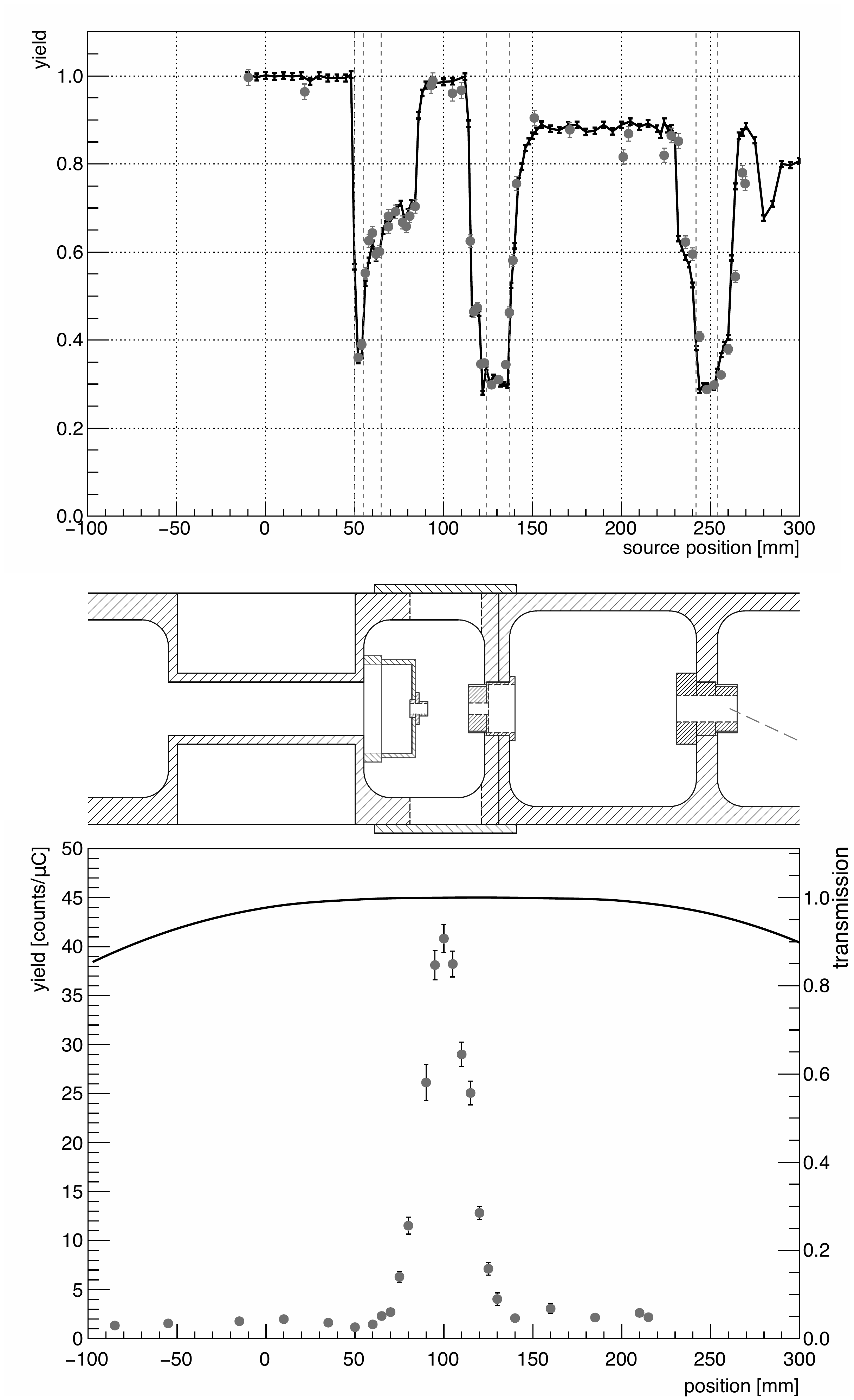}}
\caption{Top panel: target chamber's walls absorption, experimental data (filled circles) are compared with the predictions of a Geant4 simulation (dots). Both measurements and simulation are scaled to unity at position $\sim0\,$mm. Middle panel: detail of the target chamber top view. Bottom panel: gas density profile of the extended $\elem{He}{4}$ target determined through the $\elem{Li}{7}(\alpha,\alpha')$ reaction. The points are corrected for the absorption of the target chamber, according to top panel. The black line is the calculated transmission of recoils, in a selected charge state, to the end detector. The error bars shown in both panels accounts for counting statistics only.}
\label{fig:7Besource}
\end{center}
\end{figure}

\noindent The distribution of the He gas within the target cell was determined through the measurement of the yield of the broad resonance, $\Gamma_{\rm c.m.}=130$\,keV, in $\elem{Li}{7}(\alpha,\gamma\alpha')\elem{Li}{7}$ at the energy of $E_{\rm lab}=3325$\,keV, in a similar way as reported in \cite{Schuermann2013}.  In order to correct the observed $\gamma$-ray yield for the absorption by the chamber walls, the experimental setup was simulated with Geant4 \cite{Geant4}. The simulation was validated against a measurement of the relative attenuation of an uncalibrated $\elem{Be}{7}$ source that could be moved along the beam axis of the target chamber. A comparison of the experimental data with the predictions of the Geant4 simulation is shown in Fig. \ref{fig:7Besource}.

\noindent The tails of the profile are well determined and account for about 25\% of the total target thickness. The fact that a significant portion of the target gas is located outside the central cell is not an issue with respect to the separator acceptance if the yield of narrow resonances is to be measured, since beam energy can be adjusted to have the reaction to take place mainly at the center of the target. 
The effect of this feature on the measurement of non resonant cross section is discussed in Sec. \ref{sec:acceptance}

\subsection{\maybebm{\elem{F}{19}} charge state probability}

\begin{figure}[!hbtp]
\begin{center}
\resizebox{.9\hsize}{!}{\includegraphics{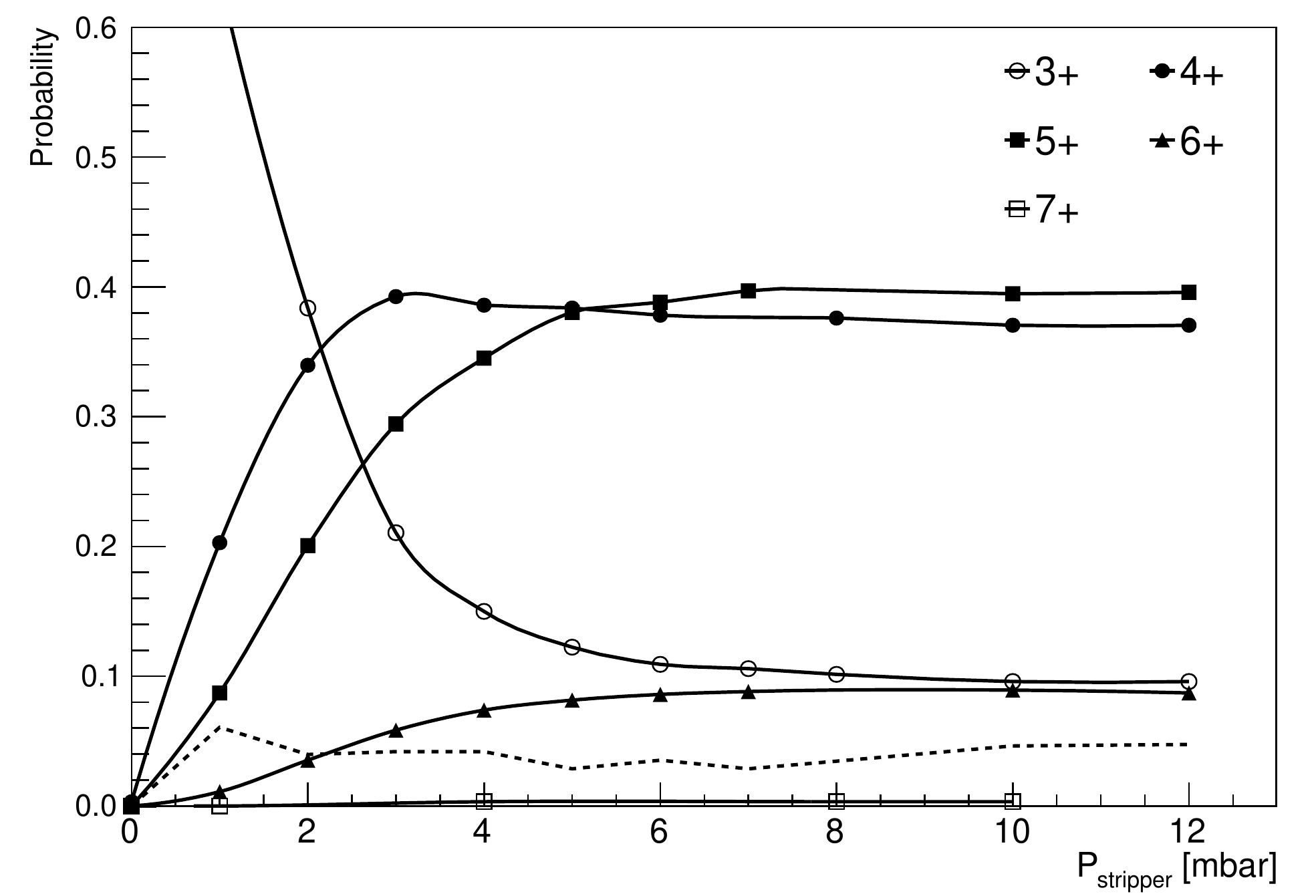}}
\caption{Charge state probability of $\elem{F}{19}$ ions as a function of the Ar post-stripper inlet pressure $P_{\rm stripper}$ at 5.0\,MeV beam energy. Lines connecting the points are to guide the eye only. The dotted line represents the unmeasured current at this energy, due to non accessible 1+, 2+ charge states and further charge exchanging in the CSSM chamber, see text for details.}
\label{fig:stripper}
\end{center}
\end{figure}

The Ar post-stripper equilibrium thickness for $\elem{F}{19}$ ions was determined through a measurement of the charge state probabilities, at several energies as a function of the stripper inlet pressure $P_{\rm stripper}$. In Fig. \ref{fig:stripper} the charge state probability as a function of the post-stripper inlet pressure is shown for the case of of 5\,MeV $^{19}\rm F^{3+}$ ions. On the basis of this measurement the working pressure of $P_{\rm stripper}=10\rm\,mbar$ was chosen.

We have also measured the charge state probabilities $\Phi_q$ of $\elem{F}{19}$ as a function of ion speed.
\begin{figure}[!hbtp]
\begin{center}
\resizebox{.9\hsize}{!}{\includegraphics{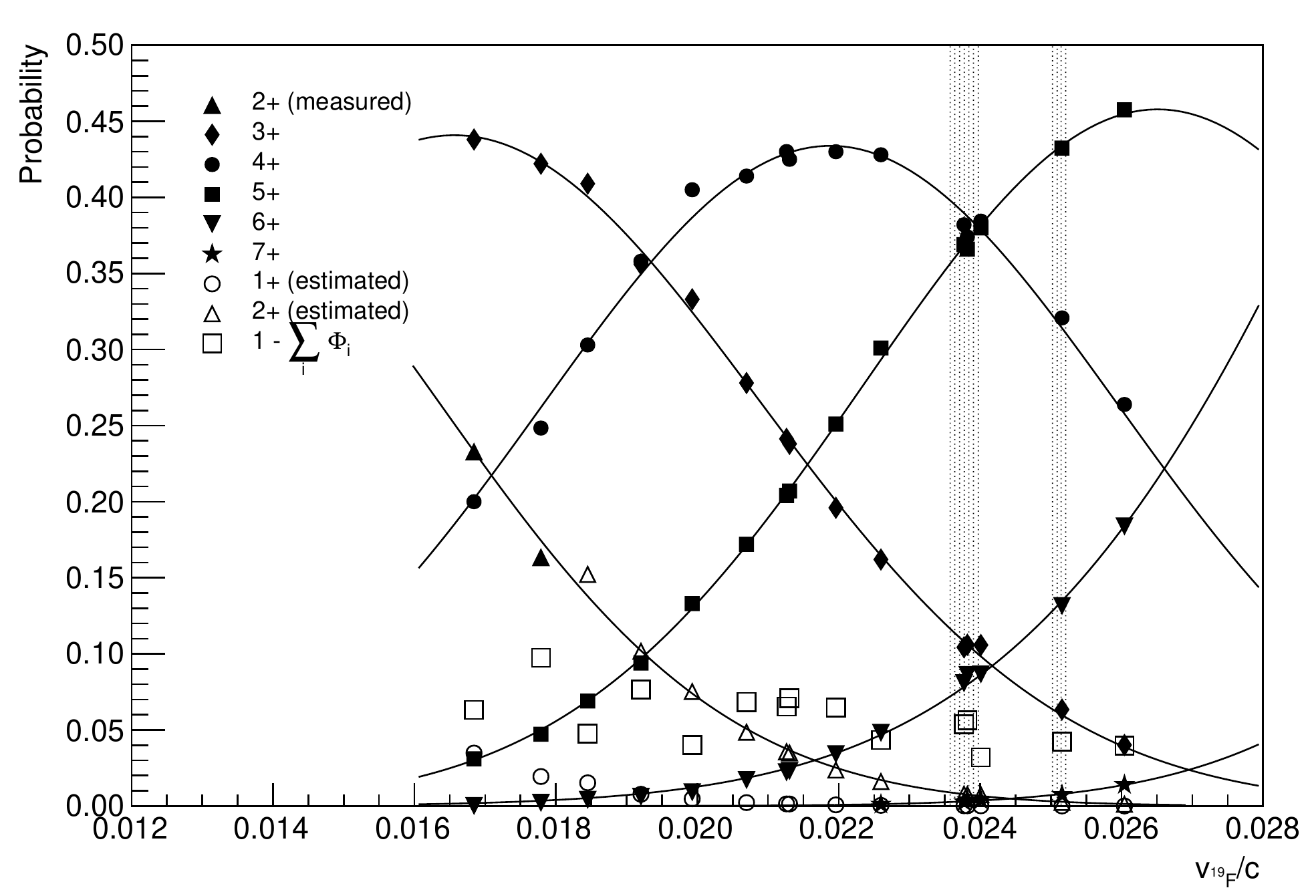}}
\caption{Charge state probability of $\elem{F}{19}$ ions emerging from the target as a function of velocity. Filled symbols are experimentally determined values, while empty symbols are estimated values for the unmeasured 1+, 2+ charge states and for the further charge exchanging in the CSSM chamber, see text for details. Curves through points are uncorrelated Gaussian fits. Vertical shaded areas indicate the energy intervals where cross section measurements were performed.}
\label{fig:ChargeStateProbability}
\end{center}
\end{figure}
Results are shown in Fig. \ref{fig:ChargeStateProbability}, the curves through the points are gaussian fits to the data, performed independently for each charge state. Due to the limitations of the CSSM magnetic field, not all of the charge states could be measured at all energies. In these cases, the unmeasured charge state probabilities, namely of 1+ and 2+, were estimated from the measured ones. In fact at a given energy, provided that the neutral and fully stripped states are negligibly populated, the probability as a function of the charge state can be assumed to be gaussian.

\noindent It has to be noted that the derived $\Phi_q$ do not correspond exactly to the charge state probabilities at the exit of the post-stripper. More precisely they are the fraction of ions that enter the triplet in the given charge state. A small difference is introduced by recoils further charge exchanging in the CSSM, where some residual Ar gas is present over a relatively long distance, leading to the loss of the ions. This feature has been verified observing the variation of the beam current after the CSSM while injecting Ar gas in the CSSM chamber only. The difference amounts altogether to about 5\%, in fact summing all the $\Phi_q$ a value of about 95\% is obtained at all energies, see Fig. \ref{fig:ChargeStateProbability}. 

\subsection{Acceptance}
\label{sec:acceptance}

\begin{figure}[!b]
\begin{center}
\resizebox{.9\hsize}{!}{\includegraphics{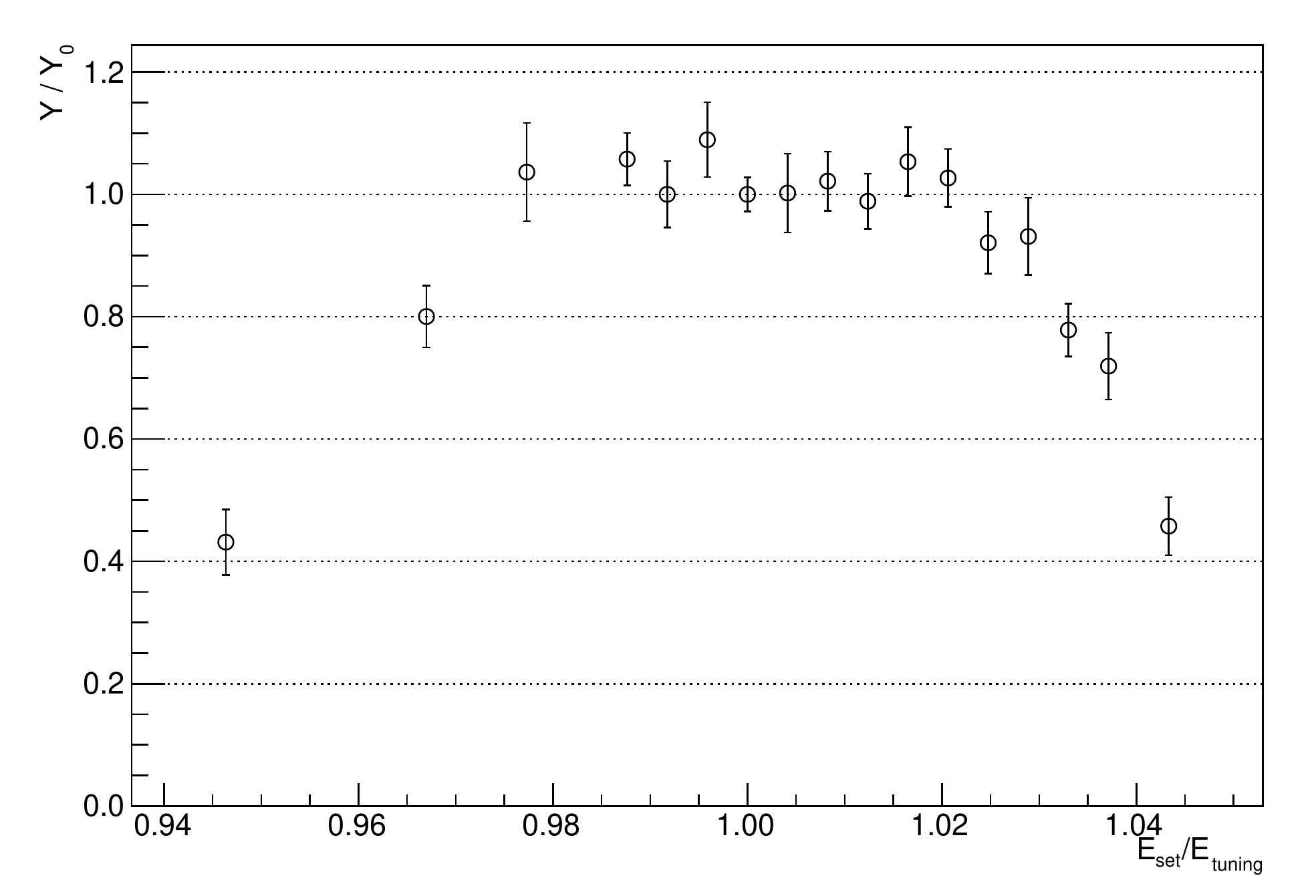}}
\caption{Ratio of the observed yield $Y$ with respect to the central yield $Y_0$ of the $E_{\rm c.m.}=1323$\,keV resonance as a function of the energy set for the separator.}
\label{fig:acceptancescan}
\end{center}
\end{figure}
The transmission of the recoils to the end detector, $T_{RMS}$, was measured to be 100\% using a $\rm^{19}F$ ion beam varying the energy and angle to scan the  volume of the phase space occupied by the recoils. An electrostatic deflection unit has been used to mimic the recoil cone with a maximum opening angle $\vartheta_{\rm max}$, which is calculated according to reaction kinematics and straggling effects due to the interaction with target and post-stripper gas.

As a further test of the separator acceptance, we have used the yield of the $E_{\rm c.m.}=1323$\,keV resonance. A scan of the target was performed and then the energy of the beam was set to  the middle of the plateau. Then several measurement were performed varying the energy to which the separator was {\em tuned}. Results are shown in Fig. \ref{fig:acceptancescan}. The experimental points show a flat-top plateau, indicating a broad region of full acceptance, and then the reaction yield sharply drops, indicating that the limit of the energy acceptance, or the limit of angular acceptance, or both, is reached.

Moreover reaction yield measurements of the $1323\,$keV resonance performed in the 3+ and 4+ charge states, characterised by quite different charge state probabilities, have given very consistent results, see Fig. \ref{fig:Yield} top panel.

\section{Experimental results and analysis}

\begin{figure}[!hbtp]
\begin{center}
\resizebox{.9\hsize}{!}{\includegraphics{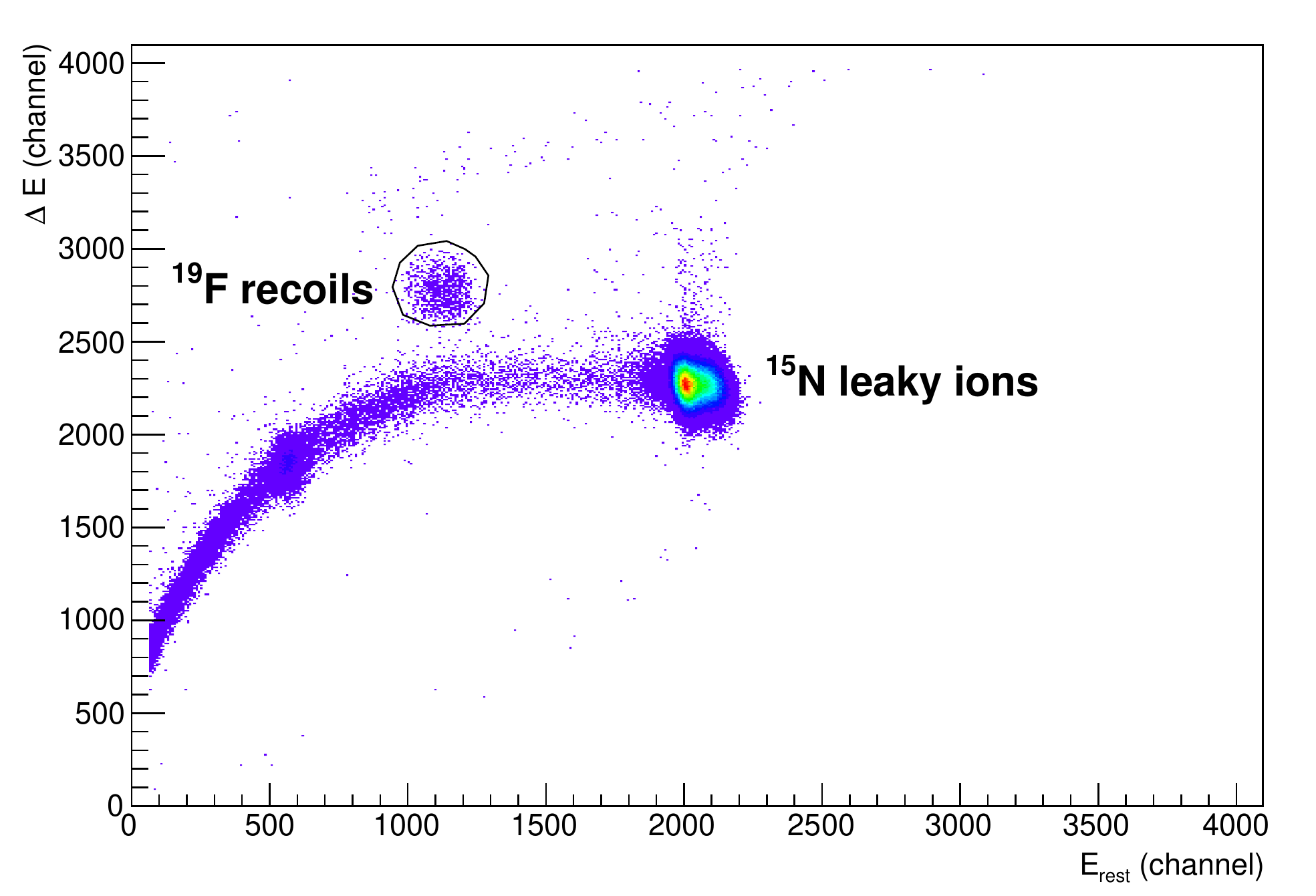}}
\caption{Sample ICT $E_{\rm rest}$-$\Delta E$ spectrum for ions identification and counting, collected at $E_{^{15}\rm N}=7.06\,$MeV.}
\label{fig:spectrum}
\end{center}
\end{figure}

\begin{figure}[!h]
\begin{center}
\resizebox{0.9\hsize}{!}{\includegraphics{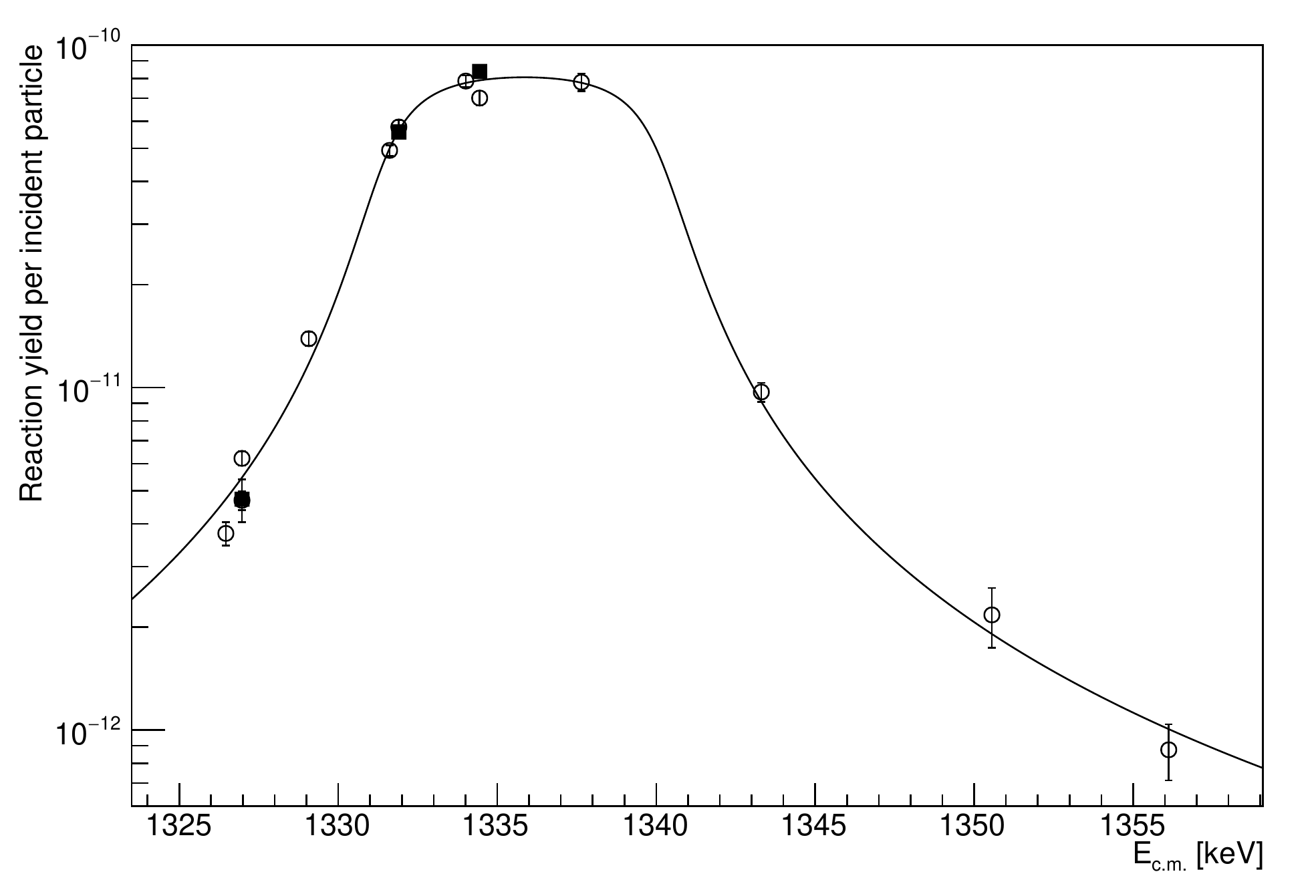}}
\resizebox{0.9\hsize}{!}{\includegraphics{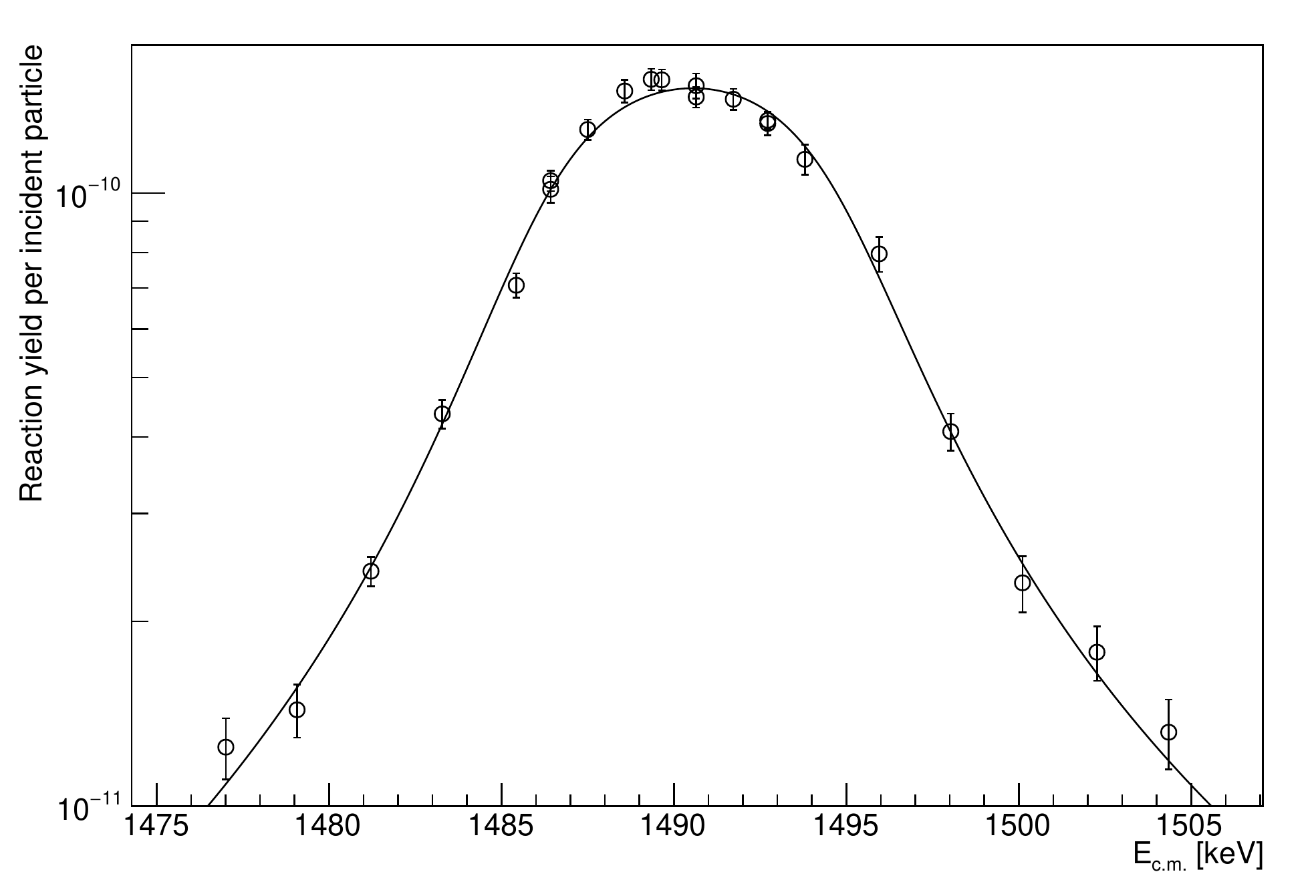}}
\caption{Reaction yield per incident projectile observed for the two broad resonances at $1323\,$keV and $1487\,$keV, top and bottom respectively. Open circles and filled squares indicate measurements of recoils in the 4+ and in the 3+ charge state, respectively.}
\label{fig:Yield}
\end{center}
\end{figure}

The reaction yield of the two broad resonances at $1323\,$keV and $1487\,$keV, corresponding to the $\rm^{19}F$ states at $E_x=5337$ and $5500.7\,$keV respectively, was measured. Ion identification and counting was done using an Ionization Chamber with a fractioned anode as a $E_{\rm rest}$-$\Delta E$ telescope (ICT). In Fig. \ref{fig:spectrum} a sample spectrum is reported. The reaction yield as a function of energy is shown in Fig. \ref{fig:Yield}.

Since both resonances are relatively broad, the expected yield has been calculated through the convolution of the resonance cross section $\sigma_{BW}(E)$ and the target profile according to Eq. \ref{eq:Yield}. The stopping power of N ions in He has a negligible variation over the target thickness, and the average value of $\rm77.2\,keV\,cm^{2}/1E18\,atoms$ is used for the analysis of both resonances. The cross section $\sigma_{BW}(E)$ is calculated using the Breit-Wigner formula
\begin{equation}
\sigma_{BW}(E) = \pi\lambdaslash^2\frac{2J+1}{(2J_t+1)(2J_p+1)}\frac{\Gamma_\alpha(E)\Gamma_\gamma(E)}{(E_R-E)^2+\left(\frac{\Gamma(E)}2\right)^2}\enspace,
\end{equation}
where $\lambdaslash$ is the projectile reduced de Broglie wavelength, $J$, $J_t$, $J_p$ are the total angular momenta of the resonance, the target nucleus and the projectile, respectively, $E_R$ is the resonance energy, and $\Gamma_\alpha$ and $\Gamma_\gamma$, are the observed partial widths. Their energy dependence is calculated according to \cite{BookIliadis}:
\begin{equation}
\Gamma_\alpha(E) = 2P_\alpha(E)\gamma_\alpha^2 \enspace,
\end{equation}
where $\gamma_\alpha^2$ is the observed reduced width and $P_\alpha(E)$ is the penetration factor
\begin{equation*}
P_\alpha(E) = R\left(\frac k{F_l^2+G_l^2}\right)\enspace,
\end{equation*}
with the radius $R=\rm5.07\,fm$, $F_l$ and $G_l$ are the regular and irregular Coulomb wave functions, respectively, while
\begin{equation}
\Gamma_\gamma(E) = \Gamma_\gamma(E_R) \sum_i B_{\gamma i}\left[ \frac{E + Q - E_{xi}}{E_R + Q - E_{xi}}\right]^{2L_i+1} \enspace,
\label{eq:GammagammaE}
\end{equation}
where $Q$ is the reaction $Q$-value, $B_{\gamma i}$ is the primary $\gamma$-ray branching ratio to the final state having excitation energy $E_{xi}$, and $L_i$ is the multipolarity of the $i$-th $\gamma$-ray transition. The $B_{\gamma i}$ values are taken from the ENSDF database \cite{ENSDF}. While the multipolarity of the primary transitions are known for the 1487\,keV resonance, they are not for the 1323\,keV resonance and are assumed to be 1. However it has to be noted that the energy dependent term of Eq. \ref{eq:GammagammaE} differs from unity at most by a fraction of a percent over the measurement energy range even in case of transitions of multipolarity 2. 

\noindent  Fits of $\sigma_{BW}(E)$, according to Eq. \ref{eq:Yield}, to the experimental data are performed using a least square function (LSF). The expected yields calculated according to our best fit values are shown in Fig. \ref{fig:Yield}.

\begin{figure}[!hbt]
\begin{center}
\resizebox{.83\hsize}{!}{\includegraphics{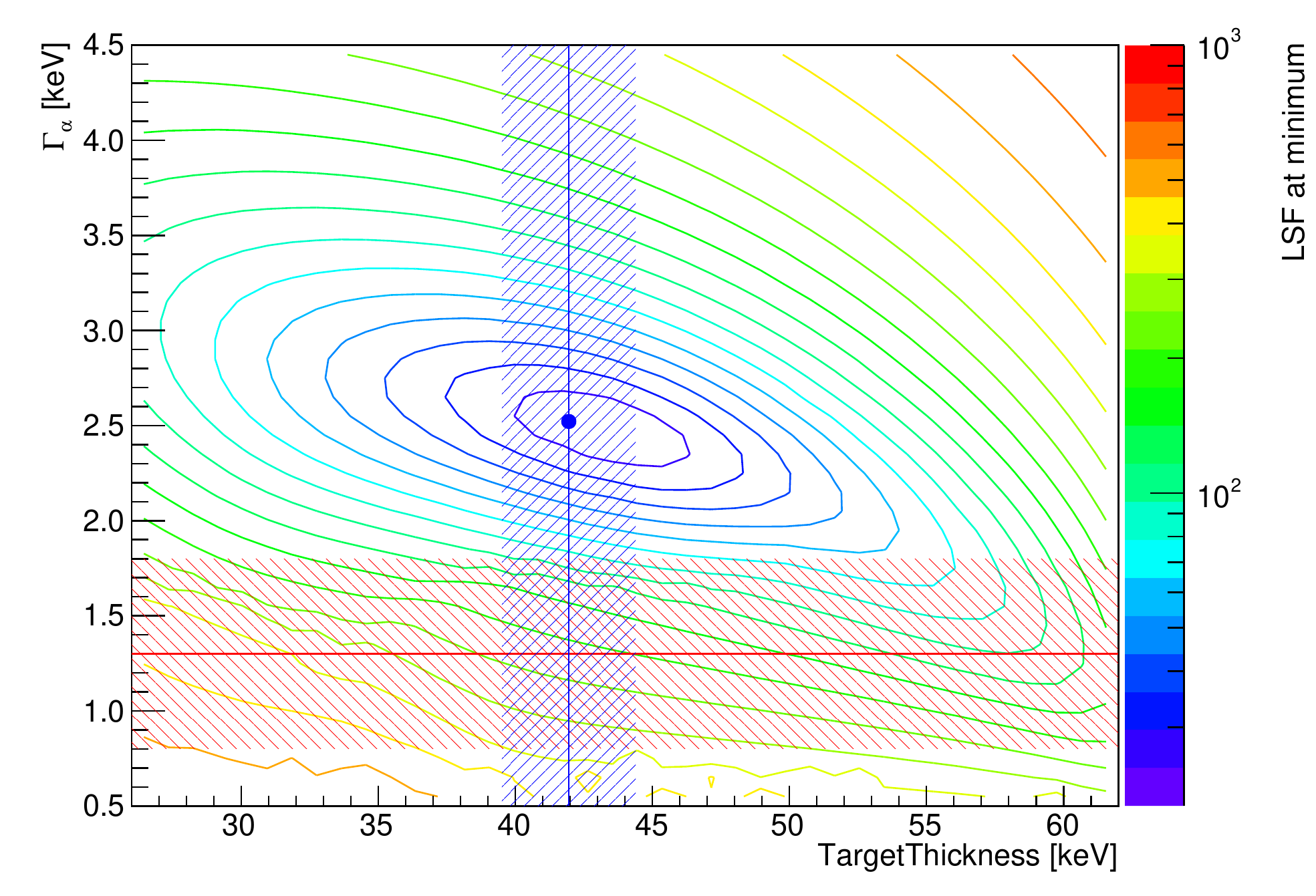}}
\resizebox{.83\hsize}{!}{\includegraphics{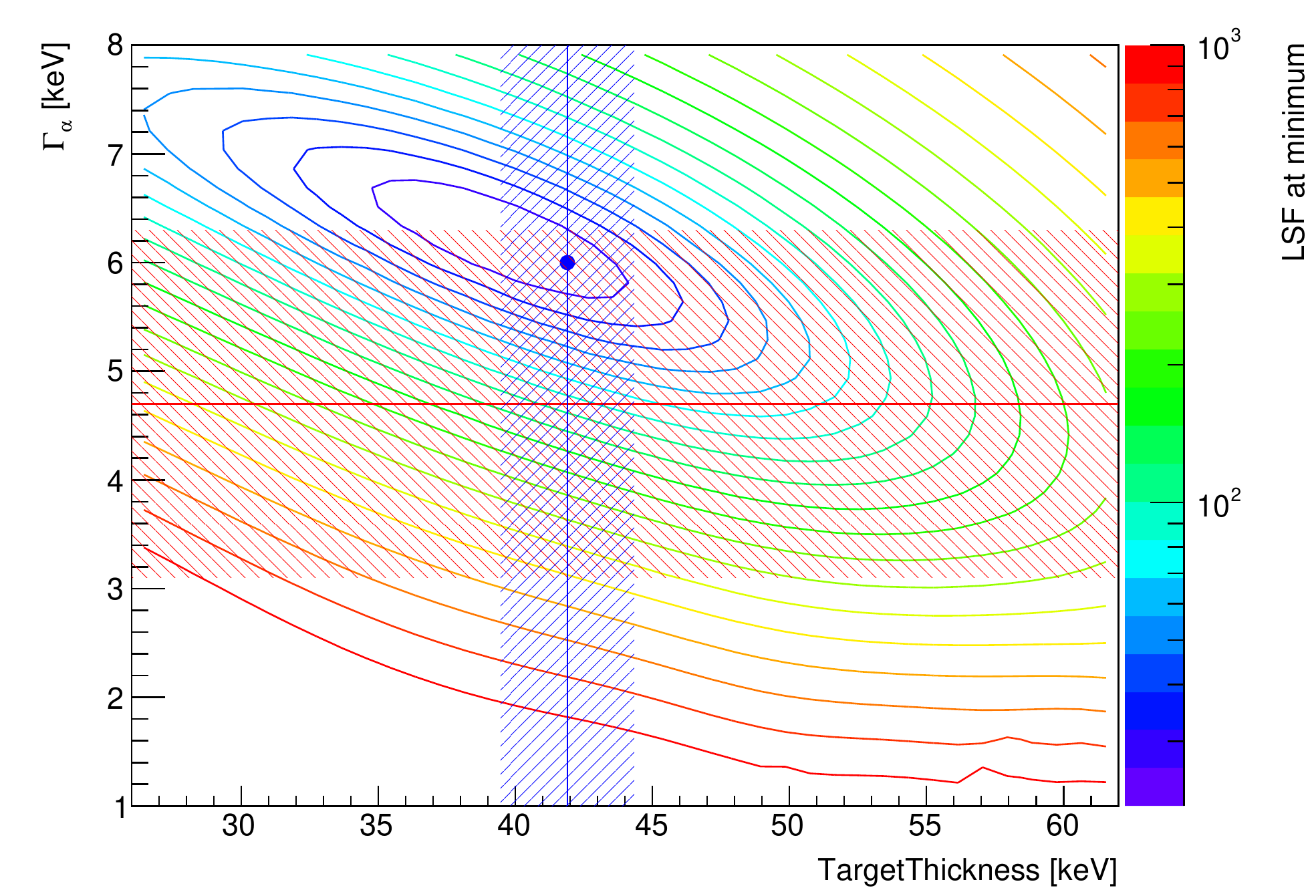}}
\caption{LSF minima contour plots, as a function of the target thickness and $\Gamma_\alpha$, for the 1.323\,MeV and the 1.487\,MeV resonances, top panel and bottom panel, respectively. The vertical line is the experimentally determined target thickness, the shaded area its uncertainty. Horizontal lines are literature values of $\Gamma_\alpha$ and shaded area their uncertainties. The dot indicates the best fit values.}
\label{fig:LSF}
\end{center}
\end{figure}

In order to exclude that this result might be an artefact of a wrong target thickness determination rather than a sizeably larger resonance total width ($\simeq\Gamma_\alpha$), a study of the correlation of these two quantities has been performed. This check was done choosing uniformly distributed random values for $T_t$ and $\Gamma_\alpha$, that were kept fixed and the LSF minimised with respect to the other parameters, namely resonance energy $E_R$ and $\Gamma_\gamma$. Results are shown, for both resonances, in Fig. \ref{fig:LSF}. Our determination of the target thickness $T_t$ leads to fit of the data with a LSF close to the absolute minimum, for both resonances, thus excluding possible issues with respect to this aspect. Literature values for $\Gamma_\alpha$ would lead to LSF minimum values quite far from the absolute minimum.

\noindent It has to be noted that even at the absolute minimum, the LSF for the 1323\,keV resonance shows quite high values (reduced $\chi^2\sim20$). Therefore for the calculation of the LSF in the fit of the 1323\,keV resonance, the statistical uncertainty of the experimental points has been inflated by a factor of 1.5\,. However this inflation has no influence on final parameter values nor on the final uncertainties estimation, since these are obtained through a Monte Carlo (MC) procedure, described below, rather than the error matrix at LSF minimum. 
\begin{table}[t]
  \begin{center}
    \begin{tabular}{lccc}
      \hline
      \hline
      & this work & \cite{Tilley1995} & \cite{Wilmes2002} \\
      \hline
      \multicolumn{2}{l}{1323\,keV resonance} \\
      $E_R$ [keV] & $1331.4\pm1.6$ & $1323\pm2$ & \\
      $\Gamma_\gamma$ [eV] & $1.62\pm0.09$ & & $1.69\pm0.14$ \\
      $\Gamma_\alpha$ [keV] & $2.51\pm0.10$ & & $1.3\pm0.5$ \\
      \hline
      \multicolumn{2}{l}{1487\,keV resonance} \\
      $E_R$ [keV] & $1486.1\pm1.9$ & $1486.7\pm1.7$ & \\
      $\Gamma_\gamma$ [eV] & $2.2\pm0.2$ & 2.13 & $1.78\pm0.17$ \\
      $\Gamma_\alpha$ [keV] & $6.0\pm0.3$ & $4\pm1$ & $4.7\pm1.6$ \\
      \hline
      \hline
    \end{tabular}
  \end{center}
  \caption{Parameters of the measured resonances as obtained from the MC procedure. Most of the uncertainty on the $E_R$ values is due to the beam energy determination.}
  \label{tab:Results}
\end{table}

\begin{figure}
\begin{center}
\resizebox{.46\hsize}{!}{\includegraphics{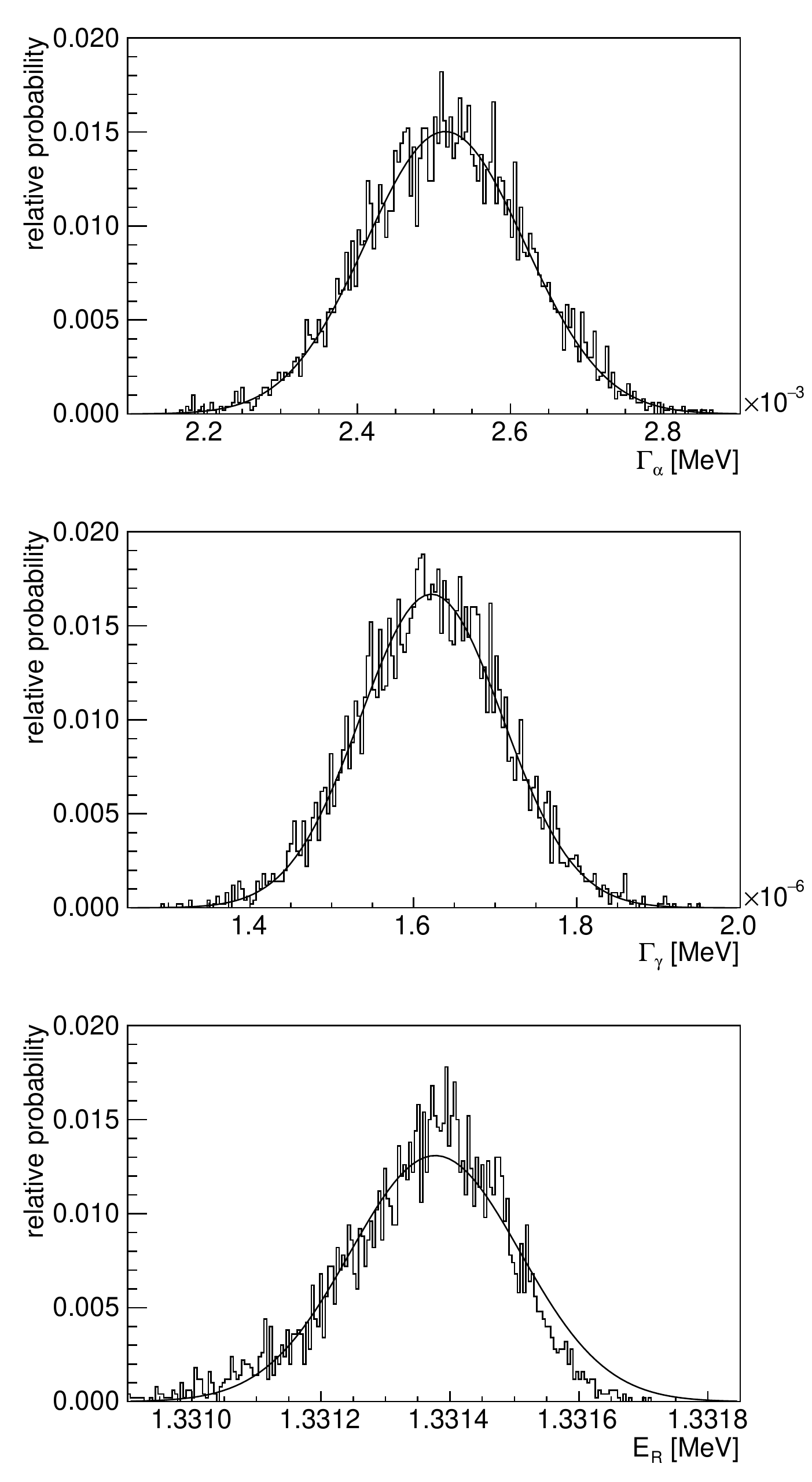}}
\resizebox{.46\hsize}{!}{\includegraphics{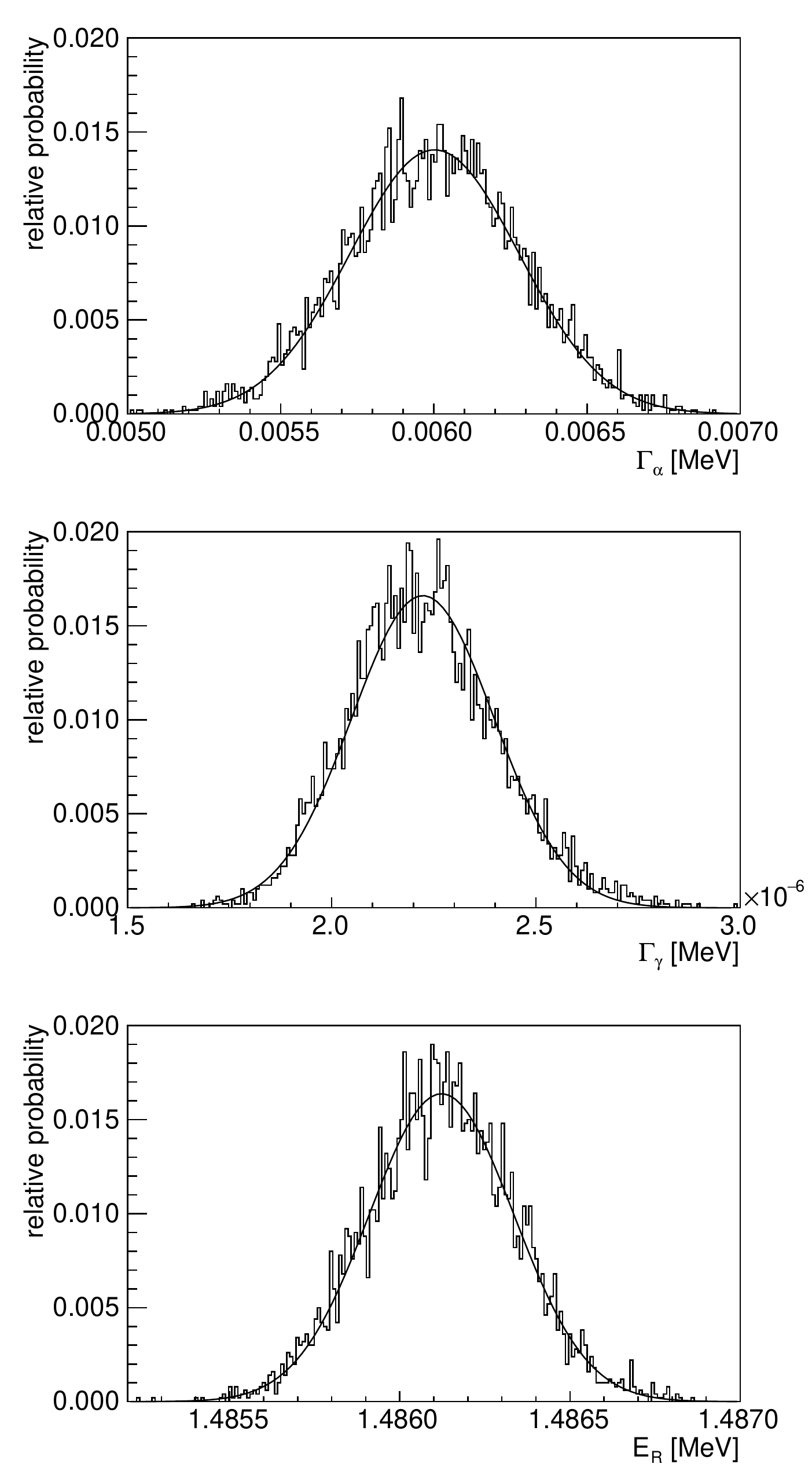}}
\caption{Parameter value distributions, $\Gamma_\alpha, \Gamma_\gamma, E_R$ (top to bottom), of the 1323\,keV (left) and the 1487\,keV (right) resonances, as obtained from the MC procedure.}
\label{fig:ParDistribution}
\end{center}
\end{figure}

The recommended values and uncertainty on the resonances parameters, reported in Table \ref{tab:Results},  are obtained through a MC procedure, so that besides the statistical uncertainty also the uncertainties on the target thickness and the other quantities contributing to the overall systematic uncertainty are correctly reflected in the results. In the MC procedure 5000 fits are performed. For each fit a pseudo-dataset is generated through a gaussian distribution of the measured values, used as central values and the uncertainty as $\sigma$, in addition the target thickness and an overall normalisation parameter are set to randomly generated values. The target thickness is generated according to a normal distribution, while the normalisation parameter is in part normally distributed, according to charge state probability, scattering rate and stopping power uncertainties,  as reported in Table \ref{tab:systematics}, and in part uniformly distributed, according to current reading uncertainty, estimated to be 3\% at all energies. Then the LSF is minimised with respect to parameters $E_R, \Gamma_\alpha$ and $\Gamma_\gamma$. The parameters distributions are shown in Fig. \ref{fig:ParDistribution}.
Some of the distributions obtained are slightly asymmetric but still rather close to normal. Therefore best value and uncertainty are obtained through a gaussian fit to the histograms, the uncertainty on beam energy determination, that contributes to $\delta E_R$, is added afterwards.
\begin{table}[!t]
  \begin{center}
    \begin{tabular}{lccc}
      \hline
      \hline
      resonance energy [keV] & $\delta \Phi_{q}$ & $\delta N_p$ & $\delta\varepsilon_{^{15}\rm N}$ \\
      \hline
      1323 & 2.1\% & 2.2\% & 5.0\% \\
      1487 & 3.2\% & 4.0\% & 5.0\% \\
      \hline
      \hline
    \end{tabular}
  \end{center}
  \caption{Relative uncertainties affecting overall normalisation: charge state probability $\delta\Phi_{q}$, number of incident projectiles $\delta N_p$, and stopping power $\delta\varepsilon_{^{15}\rm N}$. Uncertainty on current integration is 3\% at all energies.}
  \label{tab:systematics}
\end{table}

Our determination of $E_R$ for the lower energy resonance is significantly different from the literature value of 1323\,keV reported in \cite{Tilley1995}, that in turn is based on the data of \cite{Rogers1972}. It is worth noting that \cite{Tilley1995} as regards this resonance makes a reference to \cite{Kraewinkel1982}. In that work this resonance is not explicitly discussed, however the resonance profile shown, Fig. 23, panel g, appears to be consistent with a larger $E_R$ value. In addition $E_R$ values derived from $p$ and $\alpha$ inelastic scattering experiments, and $(p,\gamma)$ measurements are somewhat larger than 1323\,keV, although with larger uncertainties \cite{Tilley1995}, in better agreement with our result.

As concerns the widths, the $\Gamma_\gamma$ and $\Gamma_\alpha$ values obtained in the present work for the 1487\,keV resonance are compatible with earlier determinations \cite[and references therein]{Wilmes2002}, also the 1323\,keV resonance $\Gamma_\gamma$ is found in an excellent agreement with literature value, while a significant difference is found for the $\Gamma_\alpha$. Most notably the precision on the $\Gamma_\alpha$ values has been improved to about 5\%.

\section{Conclusions}

\begin{figure}[!htb]
\begin{center}
\resizebox{.9\hsize}{!}{\includegraphics{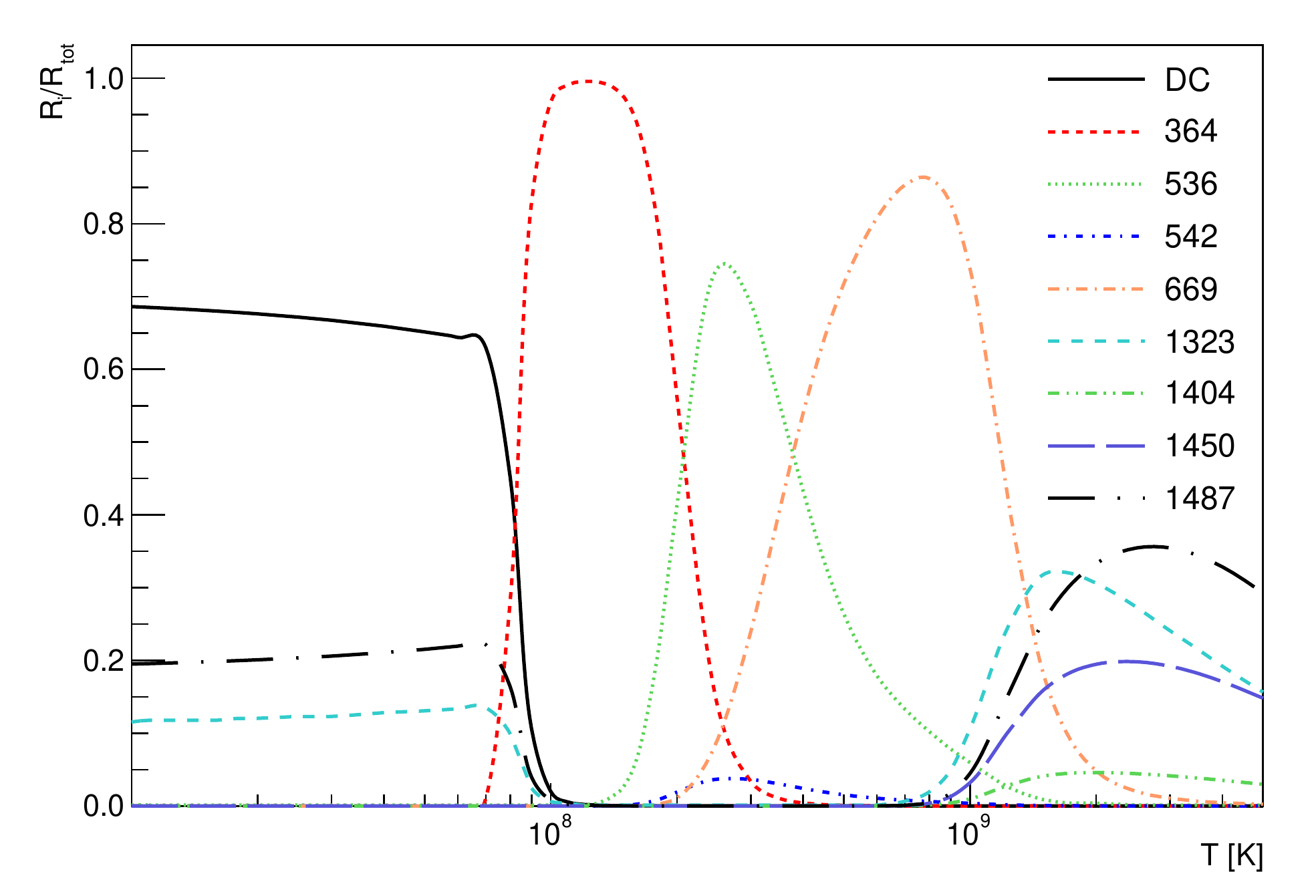}}
\caption{Fractional contribution of resonances and DC component to the total reaction rate of the $\rm^{15}N(\alpha,\gamma)^{19}F$, as a function of the temperature. The resonances are identified with their center-of-mass energy in keV.}
\label{fig:FractionalRate}
\end{center}
\end{figure}

The recoil separator ERNA has been used to directly measure the reaction yield of the two broad resonances at $E_{\rm c.m.}=1323$ and 1487\,keV. On the basis of the experimental data their $\Gamma_\gamma$ and $\Gamma_\alpha$ are determined. While agreement within uncertainty is found with earlier determination of the 1487\,keV resonance widths, a significant difference is found for the 1323\,keV $\Gamma_\alpha$. The improved determination of the broad resonances widths, influences the reaction rate, and its uncertainty, at AGB relevant temperatures. However at low  temperatures the reaction rate is dominated by the DC component and the narrow resonance at $E_{\rm c.m.}=364\,$keV. Both components are presently known only through indirect measurements \cite{deOliveira1996} and, as mentioned, are affected by large uncertainties. In Fig. \ref{fig:FractionalRate} the contribution of each resonance with respect to the total reaction rate is shown as a function of the temperature.
It is worth noting that fractional contributions to the reaction rate presented in Fig. \ref{fig:FractionalRate} are calculated according to central values and do not bring any information on the uncertainties. As mentioned the DC component and the $E_{\rm c.m.}=364\,$keV are largely uncertain, and therefore the relative contributions may vary sizeably. 

The two investigated resonances contribute to the low temperature reaction rate through their tails. Our new determination of the $\Gamma_\alpha$s increases their contribution to the reaction rate by about 15\% at relevant astrophysical energies, with respect to the rate calculated according to literature values. The relative astrophysical implications will be discussed elsewhere. 

We plan to extend the measurements towards lower energies, hopefully as far as to directly determine the strength of the $E_{\rm c.m.}=364\,$keV resonance that is presently known only through indirect measurements \cite{deOliveira1996} with a factor of 2 of uncertainty. Possibly also a direct determination of the DC component at around $E_{\rm c.m.}\sim1\,$MeV will be possible.

\acknowledgments

The Authors thank F. de Oliveira for enlightening discussions.\\*
This work was partially supported by INFN and by MIUR under the grants FIRB RBFR08549F and PRIN 20128PCN59. L.R.G. acknowledges financial support under the grants FAPESP 2014/11670-0  and Internationalization Program  UCLV 2016.
\bibliography{manuscript}

\begin{thebibliography}{29}%
\makeatletter
\providecommand \@ifxundefined [1]{%
 \@ifx{#1\undefined}
}%
\providecommand \@ifnum [1]{%
 \ifnum #1\expandafter \@firstoftwo
 \else \expandafter \@secondoftwo
 \fi
}%
\providecommand \@ifx [1]{%
 \ifx #1\expandafter \@firstoftwo
 \else \expandafter \@secondoftwo
 \fi
}%
\providecommand \natexlab [1]{#1}%
\providecommand \enquote  [1]{``#1''}%
\providecommand \bibnamefont  [1]{#1}%
\providecommand \bibfnamefont [1]{#1}%
\providecommand \citenamefont [1]{#1}%
\providecommand \href@noop [0]{\@secondoftwo}%
\providecommand \href [0]{\begingroup \@sanitize@url \@href}%
\providecommand \@href[1]{\@@startlink{#1}\@@href}%
\providecommand \@@href[1]{\endgroup#1\@@endlink}%
\providecommand \@sanitize@url [0]{\catcode `\\12\catcode `\$12\catcode
  `\&12\catcode `\#12\catcode `\^12\catcode `\_12\catcode `\%12\relax}%
\providecommand \@@startlink[1]{}%
\providecommand \@@endlink[0]{}%
\providecommand \url  [0]{\begingroup\@sanitize@url \@url }%
\providecommand \@url [1]{\endgroup\@href {#1}{\urlprefix }}%
\providecommand \urlprefix  [0]{URL }%
\providecommand \Eprint [0]{\href }%
\providecommand \doibase [0]{http://dx.doi.org/}%
\providecommand \selectlanguage [0]{\@gobble}%
\providecommand \bibinfo  [0]{\@secondoftwo}%
\providecommand \bibfield  [0]{\@secondoftwo}%
\providecommand \translation [1]{[#1]}%
\providecommand \BibitemOpen [0]{}%
\providecommand \bibitemStop [0]{}%
\providecommand \bibitemNoStop [0]{.\EOS\space}%
\providecommand \EOS [0]{\spacefactor3000\relax}%
\providecommand \BibitemShut  [1]{\csname bibitem#1\endcsname}%
\let\auto@bib@innerbib\@empty
\bibitem [{\citenamefont {{Woosley}}\ and\ \citenamefont
  {{Haxton}}(1988)}]{Woosley1988}%
  \BibitemOpen
  \bibfield  {author} {\bibinfo {author} {\bibfnamefont {S.~E.}\ \bibnamefont
  {{Woosley}}}\ and\ \bibinfo {author} {\bibfnamefont {W.~C.}\ \bibnamefont
  {{Haxton}}},\ }\href {\doibase 10.1038/334045a0} {\bibfield  {journal}
  {\bibinfo  {journal} {Nature}\ }\textbf {\bibinfo {volume} {334}},\ \bibinfo
  {pages} {45} (\bibinfo {year} {1988})}\BibitemShut {NoStop}%
\bibitem [{\citenamefont {{Meynet}}\ and\ \citenamefont
  {{Arnould}}(2000)}]{Meynet2000}%
  \BibitemOpen
  \bibfield  {author} {\bibinfo {author} {\bibfnamefont {G.}~\bibnamefont
  {{Meynet}}}\ and\ \bibinfo {author} {\bibfnamefont {M.}~\bibnamefont
  {{Arnould}}},\ }\href@noop {} {\bibfield  {journal} {\bibinfo  {journal}
  {\aap}\ }\textbf {\bibinfo {volume} {355}},\ \bibinfo {pages} {176} (\bibinfo
  {year} {2000})}\BibitemShut {NoStop}%
\bibitem [{\citenamefont {{Forestini}}\ \emph {et~al.}(1992)\citenamefont
  {{Forestini}}, \citenamefont {{Goriely}}, \citenamefont {{Jorissen}},\ and\
  \citenamefont {{Arnould}}}]{Forestini1992}%
  \BibitemOpen
  \bibfield  {author} {\bibinfo {author} {\bibfnamefont {M.}~\bibnamefont
  {{Forestini}}}, \bibinfo {author} {\bibfnamefont {S.}~\bibnamefont
  {{Goriely}}}, \bibinfo {author} {\bibfnamefont {A.}~\bibnamefont
  {{Jorissen}}}, \ and\ \bibinfo {author} {\bibfnamefont {M.}~\bibnamefont
  {{Arnould}}},\ }\href@noop {} {\bibfield  {journal} {\bibinfo  {journal}
  {\aap}\ }\textbf {\bibinfo {volume} {261}},\ \bibinfo {pages} {157} (\bibinfo
  {year} {1992})}\BibitemShut {NoStop}%
\bibitem [{\citenamefont {{Jorissen}}\ \emph {et~al.}(1992)\citenamefont
  {{Jorissen}}, \citenamefont {{Smith}},\ and\ \citenamefont
  {{Lambert}}}]{Jorissen1992}%
  \BibitemOpen
  \bibfield  {author} {\bibinfo {author} {\bibfnamefont {A.}~\bibnamefont
  {{Jorissen}}}, \bibinfo {author} {\bibfnamefont {V.~V.}\ \bibnamefont
  {{Smith}}}, \ and\ \bibinfo {author} {\bibfnamefont {D.~L.}\ \bibnamefont
  {{Lambert}}},\ }\href@noop {} {\bibfield  {journal} {\bibinfo  {journal}
  {\aap}\ }\textbf {\bibinfo {volume} {261}},\ \bibinfo {pages} {164} (\bibinfo
  {year} {1992})}\BibitemShut {NoStop}%
\bibitem [{\citenamefont {{Abia}}\ \emph {et~al.}(2009)\citenamefont {{Abia}},
  \citenamefont {{Recio-Blanco}}, \citenamefont {{de Laverny}}, \citenamefont
  {{Cristallo}}, \citenamefont {{Dom{\'{\i}}nguez}},\ and\ \citenamefont
  {{Straniero}}}]{Abia2009}%
  \BibitemOpen
  \bibfield  {author} {\bibinfo {author} {\bibfnamefont {C.}~\bibnamefont
  {{Abia}}}, \bibinfo {author} {\bibfnamefont {A.}~\bibnamefont
  {{Recio-Blanco}}}, \bibinfo {author} {\bibfnamefont {P.}~\bibnamefont {{de
  Laverny}}}, \bibinfo {author} {\bibfnamefont {S.}~\bibnamefont
  {{Cristallo}}}, \bibinfo {author} {\bibfnamefont {I.}~\bibnamefont
  {{Dom{\'{\i}}nguez}}}, \ and\ \bibinfo {author} {\bibfnamefont
  {O.}~\bibnamefont {{Straniero}}},\ }\href {\doibase
  10.1088/0004-637X/694/2/971} {\bibfield  {journal} {\bibinfo  {journal}
  {\apj}\ }\textbf {\bibinfo {volume} {694}},\ \bibinfo {pages} {971} (\bibinfo
  {year} {2009})}\BibitemShut {NoStop}%
\bibitem [{\citenamefont {{Federman}}\ \emph {et~al.}(2005)\citenamefont
  {{Federman}}, \citenamefont {{Sheffer}}, \citenamefont {{Lambert}},\ and\
  \citenamefont {{Smith}}}]{Federman2005}%
  \BibitemOpen
  \bibfield  {author} {\bibinfo {author} {\bibfnamefont {S.~R.}\ \bibnamefont
  {{Federman}}}, \bibinfo {author} {\bibfnamefont {Y.}~\bibnamefont
  {{Sheffer}}}, \bibinfo {author} {\bibfnamefont {D.~L.}\ \bibnamefont
  {{Lambert}}}, \ and\ \bibinfo {author} {\bibfnamefont {V.~V.}\ \bibnamefont
  {{Smith}}},\ }\href {\doibase 10.1086/426778} {\bibfield  {journal} {\bibinfo
   {journal} {\apj}\ }\textbf {\bibinfo {volume} {619}},\ \bibinfo {pages}
  {884} (\bibinfo {year} {2005})}\BibitemShut {NoStop}%
\bibitem [{\citenamefont {{Palacios}}\ \emph {et~al.}(2005)\citenamefont
  {{Palacios}}, \citenamefont {{Arnould}},\ and\ \citenamefont
  {{Meynet}}}]{Palacios2005}%
  \BibitemOpen
  \bibfield  {author} {\bibinfo {author} {\bibfnamefont {A.}~\bibnamefont
  {{Palacios}}}, \bibinfo {author} {\bibfnamefont {M.}~\bibnamefont
  {{Arnould}}}, \ and\ \bibinfo {author} {\bibfnamefont {G.}~\bibnamefont
  {{Meynet}}},\ }\href {\doibase 10.1051/0004-6361:20053323} {\bibfield
  {journal} {\bibinfo  {journal} {\aap}\ }\textbf {\bibinfo {volume} {443}},\
  \bibinfo {pages} {243} (\bibinfo {year} {2005})}\BibitemShut {NoStop}%
\bibitem [{\citenamefont {{Straniero}}\ \emph {et~al.}(1995)\citenamefont
  {{Straniero}}, \citenamefont {{Gallino}}, \citenamefont {{Busso}},
  \citenamefont {{Chiefei}}, \citenamefont {{Raiteri}}, \citenamefont
  {{Limongi}},\ and\ \citenamefont {{Salaris}}}]{Straniero1995}%
  \BibitemOpen
  \bibfield  {author} {\bibinfo {author} {\bibfnamefont {O.}~\bibnamefont
  {{Straniero}}}, \bibinfo {author} {\bibfnamefont {R.}~\bibnamefont
  {{Gallino}}}, \bibinfo {author} {\bibfnamefont {M.}~\bibnamefont {{Busso}}},
  \bibinfo {author} {\bibfnamefont {A.}~\bibnamefont {{Chiefei}}}, \bibinfo
  {author} {\bibfnamefont {C.~M.}\ \bibnamefont {{Raiteri}}}, \bibinfo {author}
  {\bibfnamefont {M.}~\bibnamefont {{Limongi}}}, \ and\ \bibinfo {author}
  {\bibfnamefont {M.}~\bibnamefont {{Salaris}}},\ }\href {\doibase
  10.1086/187767} {\bibfield  {journal} {\bibinfo  {journal} {\apjl}\ }\textbf
  {\bibinfo {volume} {440}},\ \bibinfo {pages} {L85} (\bibinfo {year}
  {1995})}\BibitemShut {NoStop}%
\bibitem [{\citenamefont {{Imbriani}}\ \emph {et~al.}(2012)\citenamefont
  {{Imbriani}}, \citenamefont {{deBoer}}, \citenamefont {{Best}}, \citenamefont
  {{Couder}}, \citenamefont {{Gervino}}, \citenamefont {{G{\"o}rres}},
  \citenamefont {{LeBlanc}}, \citenamefont {{Leiste}}, \citenamefont {{Lemut}},
  \citenamefont {{Stech}}, \citenamefont {{Strieder}}, \citenamefont
  {{Uberseder}},\ and\ \citenamefont {{Wiescher}}}]{Imbriani2012}%
  \BibitemOpen
  \bibfield  {author} {\bibinfo {author} {\bibfnamefont {G.}~\bibnamefont
  {{Imbriani}}}, \bibinfo {author} {\bibfnamefont {R.~J.}\ \bibnamefont
  {{deBoer}}}, \bibinfo {author} {\bibfnamefont {A.}~\bibnamefont {{Best}}},
  \bibinfo {author} {\bibfnamefont {M.}~\bibnamefont {{Couder}}}, \bibinfo
  {author} {\bibfnamefont {G.}~\bibnamefont {{Gervino}}}, \bibinfo {author}
  {\bibfnamefont {J.}~\bibnamefont {{G{\"o}rres}}}, \bibinfo {author}
  {\bibfnamefont {P.~J.}\ \bibnamefont {{LeBlanc}}}, \bibinfo {author}
  {\bibfnamefont {H.}~\bibnamefont {{Leiste}}}, \bibinfo {author}
  {\bibfnamefont {A.}~\bibnamefont {{Lemut}}}, \bibinfo {author} {\bibfnamefont
  {E.}~\bibnamefont {{Stech}}}, \bibinfo {author} {\bibfnamefont
  {F.}~\bibnamefont {{Strieder}}}, \bibinfo {author} {\bibfnamefont
  {E.}~\bibnamefont {{Uberseder}}}, \ and\ \bibinfo {author} {\bibfnamefont
  {M.}~\bibnamefont {{Wiescher}}},\ }\href {\doibase
  10.1103/PhysRevC.85.065810} {\bibfield  {journal} {\bibinfo  {journal}
  {\prc}\ }\textbf {\bibinfo {volume} {85}},\ \bibinfo {eid} {065810} (\bibinfo
  {year} {2012})}\BibitemShut {NoStop}%
\bibitem [{\citenamefont {{Lombardo}}\ \emph {et~al.}(2015)\citenamefont
  {{Lombardo}}, \citenamefont {{Dell'Aquila}}, \citenamefont {{Di Leva}},
  \citenamefont {{Indelicato}}, \citenamefont {{La Cognata}}, \citenamefont
  {{La Commara}}, \citenamefont {{Ordine}}, \citenamefont {{Rigato}},
  \citenamefont {{Romoli}}, \citenamefont {{Rosato}}, \citenamefont
  {{Spadaccini}}, \citenamefont {{Spitaleri}}, \citenamefont {{Tumino}},\ and\
  \citenamefont {{Vigilante}}}]{Lombardo2015}%
  \BibitemOpen
  \bibfield  {author} {\bibinfo {author} {\bibfnamefont {I.}~\bibnamefont
  {{Lombardo}}}, \bibinfo {author} {\bibfnamefont {D.}~\bibnamefont
  {{Dell'Aquila}}}, \bibinfo {author} {\bibfnamefont {A.}~\bibnamefont {{Di
  Leva}}}, \bibinfo {author} {\bibfnamefont {I.}~\bibnamefont {{Indelicato}}},
  \bibinfo {author} {\bibfnamefont {M.}~\bibnamefont {{La Cognata}}}, \bibinfo
  {author} {\bibfnamefont {M.}~\bibnamefont {{La Commara}}}, \bibinfo {author}
  {\bibfnamefont {A.}~\bibnamefont {{Ordine}}}, \bibinfo {author}
  {\bibfnamefont {V.}~\bibnamefont {{Rigato}}}, \bibinfo {author}
  {\bibfnamefont {M.}~\bibnamefont {{Romoli}}}, \bibinfo {author}
  {\bibfnamefont {E.}~\bibnamefont {{Rosato}}}, \bibinfo {author}
  {\bibfnamefont {G.}~\bibnamefont {{Spadaccini}}}, \bibinfo {author}
  {\bibfnamefont {C.}~\bibnamefont {{Spitaleri}}}, \bibinfo {author}
  {\bibfnamefont {A.}~\bibnamefont {{Tumino}}}, \ and\ \bibinfo {author}
  {\bibfnamefont {M.}~\bibnamefont {{Vigilante}}},\ }\href {\doibase
  10.1016/j.physletb.2015.06.073} {\bibfield  {journal} {\bibinfo  {journal}
  {\plb}\ }\textbf {\bibinfo {volume} {748}},\ \bibinfo {pages} {178} (\bibinfo
  {year} {2015})}\BibitemShut {NoStop}%
\bibitem [{\citenamefont {{Cristallo}}\ \emph {et~al.}(2014)\citenamefont
  {{Cristallo}}, \citenamefont {{Di Leva}}, \citenamefont {{Imbriani}},
  \citenamefont {{Piersanti}}, \citenamefont {{Abia}}, \citenamefont
  {{Gialanella}},\ and\ \citenamefont {{Straniero}}}]{Cristallo2014}%
  \BibitemOpen
  \bibfield  {author} {\bibinfo {author} {\bibfnamefont {S.}~\bibnamefont
  {{Cristallo}}}, \bibinfo {author} {\bibfnamefont {A.}~\bibnamefont {{Di
  Leva}}}, \bibinfo {author} {\bibfnamefont {G.}~\bibnamefont {{Imbriani}}},
  \bibinfo {author} {\bibfnamefont {L.}~\bibnamefont {{Piersanti}}}, \bibinfo
  {author} {\bibfnamefont {C.}~\bibnamefont {{Abia}}}, \bibinfo {author}
  {\bibfnamefont {L.}~\bibnamefont {{Gialanella}}}, \ and\ \bibinfo {author}
  {\bibfnamefont {O.}~\bibnamefont {{Straniero}}},\ }\href {\doibase
  10.1051/0004-6361/201424370} {\bibfield  {journal} {\bibinfo  {journal}
  {\aap}\ }\textbf {\bibinfo {volume} {570}},\ \bibinfo {eid} {A46} (\bibinfo
  {year} {2014})}\BibitemShut {NoStop}%
\bibitem [{\citenamefont {de~Oliveira}\ \emph {et~al.}(1996)\citenamefont
  {de~Oliveira}, \citenamefont {Coc}, \citenamefont {Aguer}, \citenamefont
  {Angulo}, \citenamefont {Bogaert}, \citenamefont {Kiener}, \citenamefont
  {Lefebvre}, \citenamefont {Tatischeff}, \citenamefont {Thibaud},
  \citenamefont {Fortier}, \citenamefont {Maison}, \citenamefont {Rosier},
  \citenamefont {Rotbard}, \citenamefont {Vernotte}, \citenamefont {Arnould},
  \citenamefont {Jorissen},\ and\ \citenamefont {Mowlavi}}]{deOliveira1996}%
  \BibitemOpen
  \bibfield  {author} {\bibinfo {author} {\bibfnamefont {F.}~\bibnamefont
  {de~Oliveira}}, \bibinfo {author} {\bibfnamefont {A.}~\bibnamefont {Coc}},
  \bibinfo {author} {\bibfnamefont {P.}~\bibnamefont {Aguer}}, \bibinfo
  {author} {\bibfnamefont {C.}~\bibnamefont {Angulo}}, \bibinfo {author}
  {\bibfnamefont {G.}~\bibnamefont {Bogaert}}, \bibinfo {author} {\bibfnamefont
  {J.}~\bibnamefont {Kiener}}, \bibinfo {author} {\bibfnamefont
  {A.}~\bibnamefont {Lefebvre}}, \bibinfo {author} {\bibfnamefont
  {V.}~\bibnamefont {Tatischeff}}, \bibinfo {author} {\bibfnamefont {J.-P.}\
  \bibnamefont {Thibaud}}, \bibinfo {author} {\bibfnamefont {S.}~\bibnamefont
  {Fortier}}, \bibinfo {author} {\bibfnamefont {J.}~\bibnamefont {Maison}},
  \bibinfo {author} {\bibfnamefont {L.}~\bibnamefont {Rosier}}, \bibinfo
  {author} {\bibfnamefont {G.}~\bibnamefont {Rotbard}}, \bibinfo {author}
  {\bibfnamefont {J.}~\bibnamefont {Vernotte}}, \bibinfo {author}
  {\bibfnamefont {M.}~\bibnamefont {Arnould}}, \bibinfo {author} {\bibfnamefont
  {A.}~\bibnamefont {Jorissen}}, \ and\ \bibinfo {author} {\bibfnamefont
  {N.}~\bibnamefont {Mowlavi}},\ }\href {\doibase 10.1016/0375-9474(95)00455-6}
  {\bibfield  {journal} {\bibinfo  {journal} {\npa}\ }\textbf {\bibinfo
  {volume} {597}},\ \bibinfo {pages} {231 } (\bibinfo {year}
  {1996})}\BibitemShut {NoStop}%
\bibitem [{\citenamefont {{Longland}}\ \emph {et~al.}(2010)\citenamefont
  {{Longland}}, \citenamefont {{Iliadis}}, \citenamefont {{Champagne}},
  \citenamefont {{Newton}}, \citenamefont {{Ugalde}}, \citenamefont {{Coc}},\
  and\ \citenamefont {{Fitzgerald}}}]{Longland2010}%
  \BibitemOpen
  \bibfield  {author} {\bibinfo {author} {\bibfnamefont {R.}~\bibnamefont
  {{Longland}}}, \bibinfo {author} {\bibfnamefont {C.}~\bibnamefont
  {{Iliadis}}}, \bibinfo {author} {\bibfnamefont {A.~E.}\ \bibnamefont
  {{Champagne}}}, \bibinfo {author} {\bibfnamefont {J.~R.}\ \bibnamefont
  {{Newton}}}, \bibinfo {author} {\bibfnamefont {C.}~\bibnamefont {{Ugalde}}},
  \bibinfo {author} {\bibfnamefont {A.}~\bibnamefont {{Coc}}}, \ and\ \bibinfo
  {author} {\bibfnamefont {R.}~\bibnamefont {{Fitzgerald}}},\ }\href {\doibase
  10.1016/j.nuclphysa.2010.04.008} {\bibfield  {journal} {\bibinfo  {journal}
  {\npa}\ }\textbf {\bibinfo {volume} {841}},\ \bibinfo {pages} {1} (\bibinfo
  {year} {2010})}\BibitemShut {NoStop}%
\bibitem [{\citenamefont {{Di Leva}}\ \emph {et~al.}(2012)\citenamefont {{Di
  Leva}}, \citenamefont {{Pezzella}}, \citenamefont {{De Cesare}},
  \citenamefont {{D'Onofrio}}, \citenamefont {{Gialanella}}, \citenamefont
  {{Romano}}, \citenamefont {{Romoli}}, \citenamefont {{Schuermann}},
  \citenamefont {{Terrasi}},\ and\ \citenamefont {{Imbriani}}}]{DiLeva2012}%
  \BibitemOpen
  \bibfield  {author} {\bibinfo {author} {\bibfnamefont {A.}~\bibnamefont {{Di
  Leva}}}, \bibinfo {author} {\bibfnamefont {A.}~\bibnamefont {{Pezzella}}},
  \bibinfo {author} {\bibfnamefont {N.}~\bibnamefont {{De Cesare}}}, \bibinfo
  {author} {\bibfnamefont {A.}~\bibnamefont {{D'Onofrio}}}, \bibinfo {author}
  {\bibfnamefont {L.}~\bibnamefont {{Gialanella}}}, \bibinfo {author}
  {\bibfnamefont {M.}~\bibnamefont {{Romano}}}, \bibinfo {author}
  {\bibfnamefont {M.}~\bibnamefont {{Romoli}}}, \bibinfo {author}
  {\bibfnamefont {D.}~\bibnamefont {{Schuermann}}}, \bibinfo {author}
  {\bibfnamefont {F.}~\bibnamefont {{Terrasi}}}, \ and\ \bibinfo {author}
  {\bibfnamefont {G.}~\bibnamefont {{Imbriani}}},\ }\href {\doibase
  10.1016/j.nima.2012.06.037} {\bibfield  {journal} {\bibinfo  {journal}
  {\nima}\ }\textbf {\bibinfo {volume} {689}},\ \bibinfo {pages} {98} (\bibinfo
  {year} {2012})}\BibitemShut {NoStop}%
\bibitem [{\citenamefont {{Rogalla}}\ \emph {et~al.}(2003)\citenamefont
  {{Rogalla}}, \citenamefont {{Sch{\"u}rmann}}, \citenamefont {{Strieder}},
  \citenamefont {{Aliotta}}, \citenamefont {{De Cesare}}, \citenamefont {{Di
  Leva}}, \citenamefont {{Lubritto}}, \citenamefont {{D'Onofrio}},
  \citenamefont {{Gialanella}}, \citenamefont {{Imbriani}}, \citenamefont
  {{Kluge}}, \citenamefont {{Ordine}}, \citenamefont {{Roca}}, \citenamefont
  {{R{\"o}cken}}, \citenamefont {{Rolfs}}, \citenamefont {{Romano}},
  \citenamefont {{Sch{\"u}mann}}, \citenamefont {{Terrasi}},\ and\
  \citenamefont {{Trautvetter}}}]{Rogalla2003}%
  \BibitemOpen
  \bibfield  {author} {\bibinfo {author} {\bibfnamefont {D.}~\bibnamefont
  {{Rogalla}}}, \bibinfo {author} {\bibfnamefont {D.}~\bibnamefont
  {{Sch{\"u}rmann}}}, \bibinfo {author} {\bibfnamefont {F.}~\bibnamefont
  {{Strieder}}}, \bibinfo {author} {\bibfnamefont {M.}~\bibnamefont
  {{Aliotta}}}, \bibinfo {author} {\bibfnamefont {N.}~\bibnamefont {{De
  Cesare}}}, \bibinfo {author} {\bibfnamefont {A.}~\bibnamefont {{Di Leva}}},
  \bibinfo {author} {\bibfnamefont {C.}~\bibnamefont {{Lubritto}}}, \bibinfo
  {author} {\bibfnamefont {A.}~\bibnamefont {{D'Onofrio}}}, \bibinfo {author}
  {\bibfnamefont {L.}~\bibnamefont {{Gialanella}}}, \bibinfo {author}
  {\bibfnamefont {G.}~\bibnamefont {{Imbriani}}}, \bibinfo {author}
  {\bibfnamefont {J.}~\bibnamefont {{Kluge}}}, \bibinfo {author} {\bibfnamefont
  {A.}~\bibnamefont {{Ordine}}}, \bibinfo {author} {\bibfnamefont
  {V.}~\bibnamefont {{Roca}}}, \bibinfo {author} {\bibfnamefont
  {H.}~\bibnamefont {{R{\"o}cken}}}, \bibinfo {author} {\bibfnamefont
  {C.}~\bibnamefont {{Rolfs}}}, \bibinfo {author} {\bibfnamefont
  {M.}~\bibnamefont {{Romano}}}, \bibinfo {author} {\bibfnamefont
  {F.}~\bibnamefont {{Sch{\"u}mann}}}, \bibinfo {author} {\bibfnamefont
  {F.}~\bibnamefont {{Terrasi}}}, \ and\ \bibinfo {author} {\bibfnamefont
  {H.~P.}\ \bibnamefont {{Trautvetter}}},\ }\href {\doibase
  10.1016/j.nima.2003.07.001} {\bibfield  {journal} {\bibinfo  {journal}
  {\nima}\ }\textbf {\bibinfo {volume} {513}},\ \bibinfo {pages} {573}
  (\bibinfo {year} {2003})}\BibitemShut {NoStop}%
\bibitem [{\citenamefont {{Gialanella}}\ \emph {et~al.}(2004)\citenamefont
  {{Gialanella}}, \citenamefont {{Sch{\"u}rmann}}, \citenamefont {{Strieder}},
  \citenamefont {{Di Leva}}, \citenamefont {{De Cesare}}, \citenamefont
  {{D'Onofrio}}, \citenamefont {{Imbriani}}, \citenamefont {{Klug}},
  \citenamefont {{Lubritto}}, \citenamefont {{Ordine}}, \citenamefont {{Roca}},
  \citenamefont {{R{\"o}cken}}, \citenamefont {{Rolfs}}, \citenamefont
  {{Rogalla}}, \citenamefont {{Romano}}, \citenamefont {{Sch{\"u}mann}},
  \citenamefont {{Terrasi}},\ and\ \citenamefont
  {{Trautvetter}}}]{Gialanella2004}%
  \BibitemOpen
  \bibfield  {author} {\bibinfo {author} {\bibfnamefont {L.}~\bibnamefont
  {{Gialanella}}}, \bibinfo {author} {\bibfnamefont {D.}~\bibnamefont
  {{Sch{\"u}rmann}}}, \bibinfo {author} {\bibfnamefont {F.}~\bibnamefont
  {{Strieder}}}, \bibinfo {author} {\bibfnamefont {A.}~\bibnamefont {{Di
  Leva}}}, \bibinfo {author} {\bibfnamefont {N.}~\bibnamefont {{De Cesare}}},
  \bibinfo {author} {\bibfnamefont {A.}~\bibnamefont {{D'Onofrio}}}, \bibinfo
  {author} {\bibfnamefont {G.}~\bibnamefont {{Imbriani}}}, \bibinfo {author}
  {\bibfnamefont {J.}~\bibnamefont {{Klug}}}, \bibinfo {author} {\bibfnamefont
  {C.}~\bibnamefont {{Lubritto}}}, \bibinfo {author} {\bibfnamefont
  {A.}~\bibnamefont {{Ordine}}}, \bibinfo {author} {\bibfnamefont
  {V.}~\bibnamefont {{Roca}}}, \bibinfo {author} {\bibfnamefont
  {H.}~\bibnamefont {{R{\"o}cken}}}, \bibinfo {author} {\bibfnamefont
  {C.}~\bibnamefont {{Rolfs}}}, \bibinfo {author} {\bibfnamefont
  {D.}~\bibnamefont {{Rogalla}}}, \bibinfo {author} {\bibfnamefont
  {M.}~\bibnamefont {{Romano}}}, \bibinfo {author} {\bibfnamefont
  {F.}~\bibnamefont {{Sch{\"u}mann}}}, \bibinfo {author} {\bibfnamefont
  {F.}~\bibnamefont {{Terrasi}}}, \ and\ \bibinfo {author} {\bibfnamefont
  {H.~P.}\ \bibnamefont {{Trautvetter}}},\ }\href {\doibase
  10.1016/j.nima.2003.11.386} {\bibfield  {journal} {\bibinfo  {journal}
  {\nima}\ }\textbf {\bibinfo {volume} {522}},\ \bibinfo {pages} {432}
  (\bibinfo {year} {2004})}\BibitemShut {NoStop}%
\bibitem [{\citenamefont {{Sch{\"u}rmann}}\ \emph {et~al.}(2004)\citenamefont
  {{Sch{\"u}rmann}}, \citenamefont {{Strieder}}, \citenamefont {{Di Leva}},
  \citenamefont {{Gialanella}}, \citenamefont {{De Cesare}}, \citenamefont
  {{D'Onofrio}}, \citenamefont {{Imbriani}}, \citenamefont {{Klug}},
  \citenamefont {{Lubritto}}, \citenamefont {{Ordine}}, \citenamefont {{Roca}},
  \citenamefont {{Rocken}}, \citenamefont {{Rolfs}}, \citenamefont {{Rogalla}},
  \citenamefont {{Romano}}, \citenamefont {{Schumann}}, \citenamefont
  {{Terrasi}},\ and\ \citenamefont {{Trautvetter}}}]{Schuermann2004}%
  \BibitemOpen
  \bibfield  {author} {\bibinfo {author} {\bibfnamefont {D.}~\bibnamefont
  {{Sch{\"u}rmann}}}, \bibinfo {author} {\bibfnamefont {F.}~\bibnamefont
  {{Strieder}}}, \bibinfo {author} {\bibfnamefont {A.}~\bibnamefont {{Di
  Leva}}}, \bibinfo {author} {\bibfnamefont {L.}~\bibnamefont {{Gialanella}}},
  \bibinfo {author} {\bibfnamefont {N.}~\bibnamefont {{De Cesare}}}, \bibinfo
  {author} {\bibfnamefont {A.}~\bibnamefont {{D'Onofrio}}}, \bibinfo {author}
  {\bibfnamefont {G.}~\bibnamefont {{Imbriani}}}, \bibinfo {author}
  {\bibfnamefont {J.}~\bibnamefont {{Klug}}}, \bibinfo {author} {\bibfnamefont
  {C.}~\bibnamefont {{Lubritto}}}, \bibinfo {author} {\bibfnamefont
  {A.}~\bibnamefont {{Ordine}}}, \bibinfo {author} {\bibfnamefont
  {V.}~\bibnamefont {{Roca}}}, \bibinfo {author} {\bibfnamefont
  {H.}~\bibnamefont {{Rocken}}}, \bibinfo {author} {\bibfnamefont
  {C.}~\bibnamefont {{Rolfs}}}, \bibinfo {author} {\bibfnamefont
  {D.}~\bibnamefont {{Rogalla}}}, \bibinfo {author} {\bibfnamefont
  {M.}~\bibnamefont {{Romano}}}, \bibinfo {author} {\bibfnamefont
  {F.}~\bibnamefont {{Schumann}}}, \bibinfo {author} {\bibfnamefont
  {F.}~\bibnamefont {{Terrasi}}}, \ and\ \bibinfo {author} {\bibfnamefont
  {H.~P.}\ \bibnamefont {{Trautvetter}}},\ }\href {\doibase
  10.1016/j.nima.2004.05.131} {\bibfield  {journal} {\bibinfo  {journal}
  {\nima}\ }\textbf {\bibinfo {volume} {531}},\ \bibinfo {pages} {428}
  (\bibinfo {year} {2004})}\BibitemShut {NoStop}%
\bibitem [{\citenamefont {Terrasi}\ \emph {et~al.}(2007)\citenamefont
  {Terrasi}, \citenamefont {Rogalla}, \citenamefont {De~Cesare}, \citenamefont
  {D'Onofrio}, \citenamefont {Lubritto}, \citenamefont {Marzaioli},
  \citenamefont {Passariello}, \citenamefont {Rubino}, \citenamefont
  {Sabbarese}, \citenamefont {Casa}, \citenamefont {Palmieri}, \citenamefont
  {Gialanella}, \citenamefont {Imbriani}, \citenamefont {Roca}, \citenamefont
  {Romano}, \citenamefont {Sundquist},\ and\ \citenamefont
  {Loger}}]{Terrasi2007}%
  \BibitemOpen
  \bibfield  {author} {\bibinfo {author} {\bibfnamefont {F.}~\bibnamefont
  {Terrasi}}, \bibinfo {author} {\bibfnamefont {D.}~\bibnamefont {Rogalla}},
  \bibinfo {author} {\bibfnamefont {N.}~\bibnamefont {De~Cesare}}, \bibinfo
  {author} {\bibfnamefont {A.}~\bibnamefont {D'Onofrio}}, \bibinfo {author}
  {\bibfnamefont {C.}~\bibnamefont {Lubritto}}, \bibinfo {author}
  {\bibfnamefont {F.}~\bibnamefont {Marzaioli}}, \bibinfo {author}
  {\bibfnamefont {I.}~\bibnamefont {Passariello}}, \bibinfo {author}
  {\bibfnamefont {M.}~\bibnamefont {Rubino}}, \bibinfo {author} {\bibfnamefont
  {C.}~\bibnamefont {Sabbarese}}, \bibinfo {author} {\bibfnamefont
  {G.}~\bibnamefont {Casa}}, \bibinfo {author} {\bibfnamefont {A.}~\bibnamefont
  {Palmieri}}, \bibinfo {author} {\bibfnamefont {L.}~\bibnamefont
  {Gialanella}}, \bibinfo {author} {\bibfnamefont {G.}~\bibnamefont
  {Imbriani}}, \bibinfo {author} {\bibfnamefont {V.}~\bibnamefont {Roca}},
  \bibinfo {author} {\bibfnamefont {M.}~\bibnamefont {Romano}}, \bibinfo
  {author} {\bibfnamefont {M.}~\bibnamefont {Sundquist}}, \ and\ \bibinfo
  {author} {\bibfnamefont {R.}~\bibnamefont {Loger}},\ }\href {\doibase
  10.1016/j.nimb.2007.01.139} {\bibfield  {journal} {\bibinfo  {journal}
  {\nimb}\ }\textbf {\bibinfo {volume} {259}},\ \bibinfo {pages} {14 }
  (\bibinfo {year} {2007})}\BibitemShut {NoStop}%
\bibitem [{\citenamefont {{Buompane}}\ \emph {et~al.}()\citenamefont
  {{Buompane}} \emph {et~al.}}]{Buompane2016}%
  \BibitemOpen
  \bibfield  {author} {\bibinfo {author} {\bibfnamefont {R.}~\bibnamefont
  {{Buompane}}} \emph {et~al.},\ }\href@noop {} {\ }\bibinfo {note} {To be
  published}\BibitemShut {NoStop}%
\bibitem [{\citenamefont {{Sch{\"u}rmann}}\ \emph {et~al.}(2013)\citenamefont
  {{Sch{\"u}rmann}}, \citenamefont {{Di Leva}}, \citenamefont {{Gialanella}},
  \citenamefont {{De Cesare}}, \citenamefont {{De Cesare}}, \citenamefont
  {{Imbriani}}, \citenamefont {{D'Onofrio}}, \citenamefont {{Romano}},
  \citenamefont {{Romoli}},\ and\ \citenamefont {{Terrasi}}}]{Schuermann2013}%
  \BibitemOpen
  \bibfield  {author} {\bibinfo {author} {\bibfnamefont {D.}~\bibnamefont
  {{Sch{\"u}rmann}}}, \bibinfo {author} {\bibfnamefont {A.}~\bibnamefont {{Di
  Leva}}}, \bibinfo {author} {\bibfnamefont {L.}~\bibnamefont {{Gialanella}}},
  \bibinfo {author} {\bibfnamefont {M.}~\bibnamefont {{De Cesare}}}, \bibinfo
  {author} {\bibfnamefont {N.}~\bibnamefont {{De Cesare}}}, \bibinfo {author}
  {\bibfnamefont {G.}~\bibnamefont {{Imbriani}}}, \bibinfo {author}
  {\bibfnamefont {A.}~\bibnamefont {{D'Onofrio}}}, \bibinfo {author}
  {\bibfnamefont {M.}~\bibnamefont {{Romano}}}, \bibinfo {author}
  {\bibfnamefont {M.}~\bibnamefont {{Romoli}}}, \ and\ \bibinfo {author}
  {\bibfnamefont {F.}~\bibnamefont {{Terrasi}}},\ }\href {\doibase
  10.1140/epja/i2013-13080-1} {\bibfield  {journal} {\bibinfo  {journal}
  {\epja}\ }\textbf {\bibinfo {volume} {49}},\ \bibinfo {pages} {80} (\bibinfo
  {year} {2013})}\BibitemShut {NoStop}%
\bibitem [{IAE()}]{IAEAwebsite}%
  \BibitemOpen
  \href {https://www-nds.iaea.org/stopping/stopping_heav.html} {\enquote
  {\bibinfo {title} {https://www-nds.iaea.org/},}\ }\bibinfo {note} {Accessed
  October 2016}\BibitemShut {NoStop}%
\bibitem [{\citenamefont {Price}\ \emph {et~al.}(1993)\citenamefont {Price},
  \citenamefont {Simons}, \citenamefont {Stern}, \citenamefont {Land},
  \citenamefont {Guardala}, \citenamefont {Brennan},\ and\ \citenamefont
  {Stumborg}}]{Price1993}%
  \BibitemOpen
  \bibfield  {author} {\bibinfo {author} {\bibfnamefont {J.~L.}\ \bibnamefont
  {Price}}, \bibinfo {author} {\bibfnamefont {D.~G.}\ \bibnamefont {Simons}},
  \bibinfo {author} {\bibfnamefont {S.~H.}\ \bibnamefont {Stern}}, \bibinfo
  {author} {\bibfnamefont {D.~J.}\ \bibnamefont {Land}}, \bibinfo {author}
  {\bibfnamefont {N.~A.}\ \bibnamefont {Guardala}}, \bibinfo {author}
  {\bibfnamefont {J.~G.}\ \bibnamefont {Brennan}}, \ and\ \bibinfo {author}
  {\bibfnamefont {M.~F.}\ \bibnamefont {Stumborg}},\ }\href {\doibase
  10.1103/PhysRevA.47.2913} {\bibfield  {journal} {\bibinfo  {journal} {\pra}\
  }\textbf {\bibinfo {volume} {47}},\ \bibinfo {pages} {2913} (\bibinfo {year}
  {1993})}\BibitemShut {NoStop}%
\bibitem [{\citenamefont {Agostinelli}\ \emph {et~al.}(2003)\citenamefont
  {Agostinelli} \emph {et~al.}}]{Geant4}%
  \BibitemOpen
  \bibfield  {author} {\bibinfo {author} {\bibfnamefont {S.}~\bibnamefont
  {Agostinelli}} \emph {et~al.},\ }\href {\doibase
  10.1016/S0168-9002(03)01368-8} {\bibfield  {journal} {\bibinfo  {journal}
  {\nima}\ }\textbf {\bibinfo {volume} {506}},\ \bibinfo {pages} {250}
  (\bibinfo {year} {2003})}\BibitemShut {NoStop}%
\bibitem [{\citenamefont {{Iliadis}}(2007)}]{BookIliadis}%
  \BibitemOpen
  \bibfield  {author} {\bibinfo {author} {\bibfnamefont {C.}~\bibnamefont
  {{Iliadis}}},\ }\href@noop {} {\emph {\bibinfo {title} {Nuclear Physics of
  Stars}}}\ (\bibinfo  {publisher} {Wiley-VCH Verlag, Wenheim, Germany},\
  \bibinfo {year} {2007})\BibitemShut {NoStop}%
\bibitem [{{ENSDF}()}]{ENSDF}%
  \BibitemOpen
  {ENSDF},\ \href {http://www.nndc.bnl.gov/ensdf/} {\enquote {\bibinfo {title}
  {Evaluated and compiled nuclear structure data},}\ }\bibinfo {note} {Accessed
  September 2016}\BibitemShut {NoStop}%
\bibitem [{\citenamefont {Tilley}\ \emph {et~al.}(1995)\citenamefont {Tilley},
  \citenamefont {Weller}, \citenamefont {Cheves},\ and\ \citenamefont
  {Chasteler}}]{Tilley1995}%
  \BibitemOpen
  \bibfield  {author} {\bibinfo {author} {\bibfnamefont {D.~R.}\ \bibnamefont
  {Tilley}}, \bibinfo {author} {\bibfnamefont {H.~R.}\ \bibnamefont {Weller}},
  \bibinfo {author} {\bibfnamefont {C.~M.}\ \bibnamefont {Cheves}}, \ and\
  \bibinfo {author} {\bibfnamefont {R.~M.}\ \bibnamefont {Chasteler}},\ }\href
  {\doibase 10.1016/0375-9474(95)00338-1} {\bibfield  {journal} {\bibinfo
  {journal} {\np}\ }\textbf {\bibinfo {volume} {595}},\ \bibinfo {pages} {1}
  (\bibinfo {year} {1995})}\BibitemShut {NoStop}%
\bibitem [{\citenamefont {{Wilmes}}\ \emph {et~al.}(2002)\citenamefont
  {{Wilmes}}, \citenamefont {{Wilmes}}, \citenamefont {{Staudt}}, \citenamefont
  {{Mohr}},\ and\ \citenamefont {{Hammer}}}]{Wilmes2002}%
  \BibitemOpen
  \bibfield  {author} {\bibinfo {author} {\bibfnamefont {S.}~\bibnamefont
  {{Wilmes}}}, \bibinfo {author} {\bibfnamefont {V.}~\bibnamefont {{Wilmes}}},
  \bibinfo {author} {\bibfnamefont {G.}~\bibnamefont {{Staudt}}}, \bibinfo
  {author} {\bibfnamefont {P.}~\bibnamefont {{Mohr}}}, \ and\ \bibinfo {author}
  {\bibfnamefont {J.~W.}\ \bibnamefont {{Hammer}}},\ }\href {\doibase
  10.1103/PhysRevC.66.065802} {\bibfield  {journal} {\bibinfo  {journal}
  {\prc}\ }\textbf {\bibinfo {volume} {66}},\ \bibinfo {eid} {065802} (\bibinfo
  {year} {2002})}\BibitemShut {NoStop}%
\bibitem [{\citenamefont {Rogers}\ \emph {et~al.}(1972)\citenamefont {Rogers},
  \citenamefont {Beukens},\ and\ \citenamefont {Diamond}}]{Rogers1972}%
  \BibitemOpen
  \bibfield  {author} {\bibinfo {author} {\bibfnamefont {D.~W.~O.}\
  \bibnamefont {Rogers}}, \bibinfo {author} {\bibfnamefont {R.~P.}\
  \bibnamefont {Beukens}}, \ and\ \bibinfo {author} {\bibfnamefont {W.~T.}\
  \bibnamefont {Diamond}},\ }\href {\doibase 10.1139/p72-322} {\bibfield
  {journal} {\bibinfo  {journal} {Canadian Journal of Physics}\ }\textbf
  {\bibinfo {volume} {50}},\ \bibinfo {pages} {2428} (\bibinfo {year}
  {1972})}\BibitemShut {NoStop}%
\bibitem [{\citenamefont {Kr{\"a}winkel}\ \emph {et~al.}(1982)\citenamefont
  {Kr{\"a}winkel} \emph {et~al.}}]{Kraewinkel1982}%
  \BibitemOpen
  \bibfield  {author} {\bibinfo {author} {\bibfnamefont {H.}~\bibnamefont
  {Kr{\"a}winkel}} \emph {et~al.},\ }\href {\doibase 10.1007/BF01421513}
  {\bibfield  {journal} {\bibinfo  {journal} {\zpa}\ }\textbf {\bibinfo
  {volume} {304}},\ \bibinfo {pages} {307} (\bibinfo {year}
  {1982})}\BibitemShut {NoStop}%
\end{thebibliography}%

\end{document}